%% file: Perf-paper.tex
\documentclass[11pt]{article}
\pdfoutput=1
\usepackage{graphicx}
\usepackage{amssymb}
\usepackage{amsmath}
\usepackage[left]{lineno}
\usepackage{booktabs}
\usepackage{multirow}
\usepackage[title]{appendix}
\usepackage{lineno}
\usepackage{xcolor} 
\usepackage{subcaption} 
\usepackage{cite} 
\usepackage{xspace}
\usepackage[normalem]{ulem}

\usepackage{alphalph}
\usepackage[alsoload=hep,binary-units=true]{siunitx} 
\DeclareSIUnit\clight{\text{\ensuremath{c}}}

\newcommand{\Kp}{\ensuremath{K^{+}}}
\newcommand{\kp}{\ensuremath{K^{+}}}

\newcommand{\pip}{\ensuremath{\pi^{+}}}
\newcommand{\pim}{\ensuremath{\pi^{-}}}

\newcommand{\piz}{\ensuremath{\pi^{0}}}

\newcommand{\mup}{\ensuremath{\mu^{+}}}

\newcommand{\kpipipi}{\ensuremath{\Kp \to \pip\pip\pim}}
\newcommand{\kpinunu}{\ensuremath{\Kp \to \pip\nu\bar{\nu}}}

\newcommand{\kpipi}{\ensuremath{\kp \to \pip \piz}}
\newcommand{\pgg}{\ensuremath{\piz \to \gamma \gamma}}

\newcommand{\kmunu}{\ensuremath{\kp \to \mup\nu}}

\newcommand{\trig}[1]{\ensuremath{\mathrm{#1}}}

\newcommand{\cherenkov}{Cherenkov}

\newcommand{\FV}{FV}

\newcommand{\lzCalo}{calorimeter L0 trigger}

\begin{document}
\pagenumbering{arabic}
\centerline{\LARGE EUROPEAN ORGANIZATION FOR NUCLEAR RESEARCH}
\vspace{15mm}
\begin{flushright}
CERN-EP-2022-165\\
December 22, 2022\\
 \vspace{2mm}
\end{flushright}
\vspace{15mm}

\begin{center}
\Large{\bf Performance of the NA62 trigger system \\
\vspace{5mm}
}
The NA62 Collaboration
\end{center}
\begin{abstract}
The NA62 experiment at CERN targets the measurement of the ultra-rare 
$K^{+}~\to~\pi^{+}\nu\bar{\nu}$ decay, and carries out a broad physics programme that includes probes for symmetry violations and searches for exotic particles. 
Data were collected in 2016--2018 using a multi-level trigger system, which
is described highlighting performance studies based on 2018 data.
\end{abstract}
\vspace{20mm}

\begin{center}
{\em Accepted for publication in JHEP} 
\end{center}



\clearpage
\input{run1Trigperf_2.tex}
\clearpage

\tableofcontents
\section{Introduction}
\label{sec:intro}

The NA62 experiment at CERN is devoted to precision tests of the Standard Model (SM) with \kp\ decays using a decay-in-flight technique. 
The main goal of NA62 is the measurement of the branching ratio (BR) of the ultra-rare decay \kpinunu, which is both precisely predicted by the SM and discriminating among possible SM extensions.
The branching ratio is predicted to be BR$(\kpinunu) = (8.4 \pm 1.0) \times 10^{-11}$ in the SM \cite{PnnTheory}.
The smallness of the expected BR and the presence of two undetected neutrinos in the final state make this measurement challenging.

Data collected in 2016--2018 allowed significant progress in the study of the \kpinunu ~decay.
The analysis of 2016 data demonstrated the feasibility of measuring BR(\kpinunu) using a decay-in-flight technique \cite{Pnn2016}.
With data collected in 2017, NA62 observed two SM \kpinunu\ signal candidates with an expected background of \num{1.5} events, ensuring the robustness of the measurement strategy and background suppression capabilities \cite{Pnn2017}.
The result based on data collected in 2018 provided first evidence for the \kpinunu\ decay with \num{17} signal candidates and an expected background of \num{5.3} events \cite{Pnn2018}.
The combined result
 $$\text{BR}(\kpinunu) = (10.6^{+4.0}_{-3.4} \big\vert_{\text{stat}} \pm 0.9_{\text{syst}}) \times 10^{-11}$$
gives evidence for this decay with a significance of 3.4 standard deviations. 

A large dataset of  $6 \times 10^{12}$ \kp\ decays allows several other physics goals to be pursued, including searches for decays forbidden by SM symmetries and searches for hidden-sector particles produced in \kp\ decays~\cite{Goudzovski:2022vbt}.
The NA62 experiment also operates in a beam-dump configuration, providing the opportunity to seek decays of hidden-sector particles with masses above the \kp\ mass;
the sensitivity to hidden-sector particles is reported in~\cite{HS}.
A dataset corresponding to \num{3e16} protons on target (POT) was collected in beam-dump configuration in 2016--2018.

The diverse physics programme demands a selective and flexible trigger system.
In the following, the structure and performance of the trigger system that operated in 2018 are described.

\section{The NA62 experiment}
\label{subsec:NA62detector}

\begin{figure}[t]
 \centering
\includegraphics[width=1.0\textwidth]{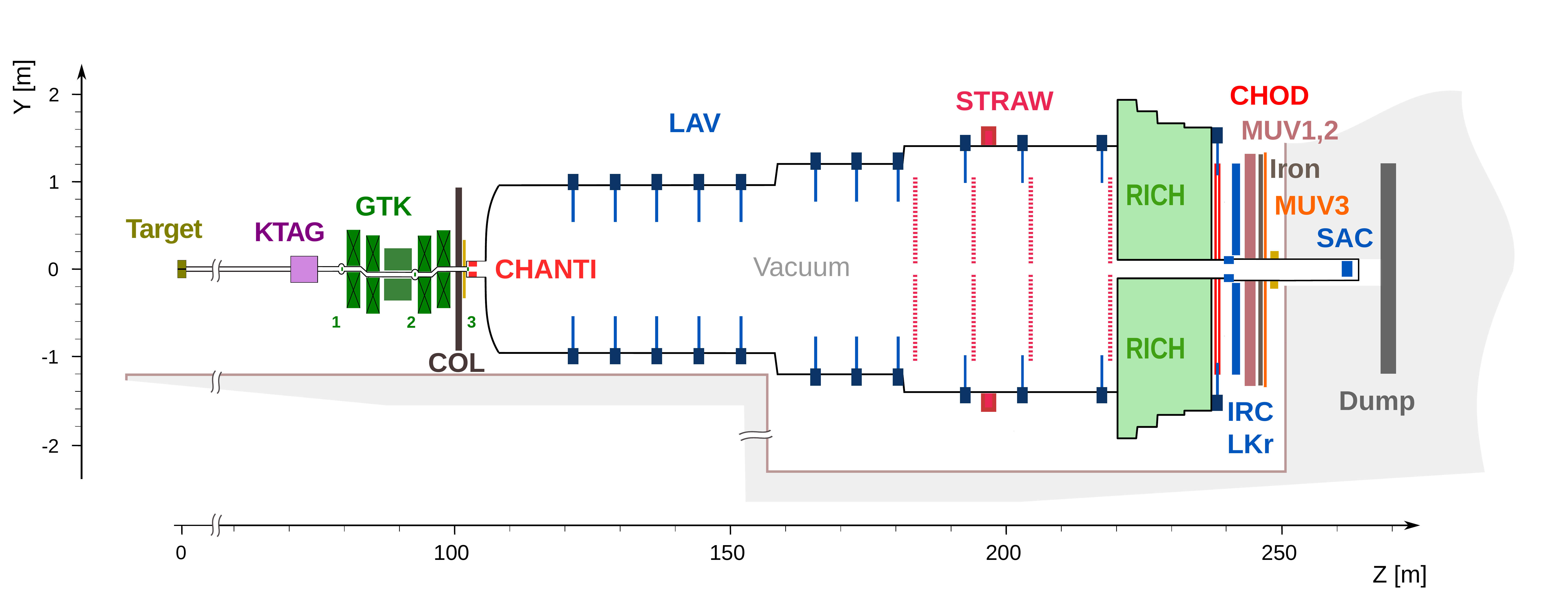}
\put (-380, 151){\makebox[0.7\textwidth][r]{\scriptsize\textsf{\textbf{\textcolor{red}{NA48-CHOD}}}}}
\caption{Schematic side view of the NA62 beamline and detector.}
 \label{fig:detector}
\end{figure}

The NA62 experiment is sketched in Fig.~\ref{fig:detector} and is described in detail in \cite{DetectorPaper}. 
A beam of positively charged hadrons is formed by impinging \SI{400}{\giga\eV/\clight} protons extracted from the CERN SPS onto a beryllium target in spills of \SI{3}{\second} effective duration. 
The target defines the origin of the NA62 reference system: the beam travels along the Z axis in the positive direction (downstream), the Y axis points vertically up, and the X axis is horizontal and directed to form a right-handed coordinate system.
The beam has a nominal momentum of \SI{75}{\giga\eV/\clight} and consists of 70\% \pip, 23\% protons and 6\% \kp.

Particles in the hadron beam are characterised by a differential Cherenkov counter (KTAG) and a three-station silicon-pixel beam spectrometer (GTK~1--3). The beam passes through a vacuum tank which contains a \SI{60}{m} long fiducial volume (FV).
Charged products of \kp\ decays inside the \FV\ travel downstream, and are detected by a straw-chamber spectrometer (STRAW), a ring-imaging \cherenkov{} detector (RICH), and 
two plastic scintillator hodoscopes:
CHOD made of one plane of tiles;
NA48-CHOD with two planes of slabs.
The RICH, filled with neon at atmospheric pressure, tags the charged products with a timing precision of better than \SI{100}{\pico\second} and provides particle identification.
Further particle identification is provided by an electromagnetic calorimeter (LKr), two hadronic calorimeters (MUV1, MUV2), and a scintillator-tile hodoscope (MUV3) located behind a \SI{80}{\centi\meter} iron wall.
Dedicated photon detectors (LAV, IRC, SAC) are placed to form a hermetic photon-veto system. 
An argon ionisation chamber (Argonion) located downstream of the SAC measures the number of particles in the hadron beam in each spill~\cite{Argonion}.

The beam intensity is defined in terms of the particle rate exiting the GTK.
The nominal beam intensity is \SI{750}{\mega\hertz}, of which \SI{45}{\mega\hertz} is \kp.
There are about \SI{10}{\mega\hertz} of \kp\ decays within the \SI{150}{\meter} region between the exit of the GTK and the MUV3 detector, with \SI{5}{\mega\hertz} occurring inside the \FV.
A muon halo originating primarily from \kp\ and \pip\ decays upstream of the \FV\ accompanies the hadron beam, and introduces an additional particle rate of several \mega\hertz\ in the detectors downstream of the \FV.
To handle the high particle rate, NA62 utilises timing measurements with sub-\SI{100}{\pico\second} precision in the RICH, KTAG, and GTK, and from \SI{100}{\pico\second} to \SI{1}{\nano\second} in the other detectors.

The intensity of the hadron beam is estimated in two ways.
The average beam intensity during the spill is evaluated using the number of beam particles observed by the Argonion detector.
This information is available immediately after the spill, with the nominal beam intensity corresponding to about $2.0\times10^{9}$ Argonion counts.
The instantaneous beam intensity is evaluated on an event-by-event basis by measuring the GTK signal rate. 
This information is available after event reconstruction. 

When operating in beam-dump configuration, the target is removed and the proton beam is stopped in movable collimators that act as a beam dump, located about \SI{45}{\meter} upstream of the KTAG. In this configuration, the particle rate in the detectors downstream of the \FV\ is \SI{300}{\kilo\hertz}, dominated by muons produced in the beam dump.

\section{The NA62 trigger system}
\label{sec:trigger}

The NA62 trigger system is arranged in two stages. 
The first stage is a low-level hardware trigger (L0) that uses custom programmable electronics to process data from a subset of detectors and form a L0 trigger decision.
The subsequent L1 trigger stage is a software trigger running on a dedicated data-acquisition (DAQ) computing cluster (DAQ-farm).
Each trigger stage is designed to reduce the event rate by roughly a factor of 10, going from about \SI{10}{\mega\hertz} of events from \kp\ decays occurring in the detector to about \SI{100}{\kilo\hertz} written to permanent storage. 

The data are collected through \textit{trigger lines}.
Each trigger line is a sequence of L0 and L1 trigger conditions designed to select a specific category of events.
Each condition can either be required or forbidden (vetoed).
Applying trigger conditions in veto introduces \textit{random veto}, whereby events are rejected by the trigger due to unrelated activity in the detector. 
An event is stored if it is accepted by at least one of the trigger lines.

Ten trigger lines are defined to serve the \kpinunu\ branching ratio measurement and cover the breadth of the \kp\ physics programme.
The \emph{PNN} line collects data for the \kpinunu\ branching ratio measurement.
The \emph{non-muon} (Non-$\mu$) line is used to collect \kp\ decays without muons in the final state. The \emph{multi-track} (MT) line collects samples of multi-track \kp\ decays.
The \emph{di-muon multi-track} (2$\mu$MT), \emph{electron multi-track} ($e$MT), and \emph{muon multi-track} ($\mu$MT) lines collect \kp\ decays to multi-lepton final states. 
The \emph{displaced muon} (DV-$\mu$) and \emph{displaced di-muon} (DV-2$\mu$) lines collect events with a vertex in the \FV\ displaced with respect to the nominal beam axis. 
The \emph{neutrino} line collects \kmunu\ events in which the neutrino interacts with the LKr material, for a proof-of-concept study of flavour-tagged neutrino interaction events.
Finally, the \emph{control} line collects minimum-bias events for calibration and data-quality monitoring, for the normalization of the \kpinunu\ branching ratio measurement, and for the measurement of detector and trigger efficiencies.
Three trigger lines are defined for use in beam-dump configuration -- \emph{neutral}, \emph{charged}, and \emph{control} -- which collect di-photon, multi-track and control data samples, respectively.

\subsection{The low-level trigger}

\begin{figure}[t]
\centering
\begin{subfigure}[b]{0.45\textwidth}
\includegraphics[width=\textwidth]{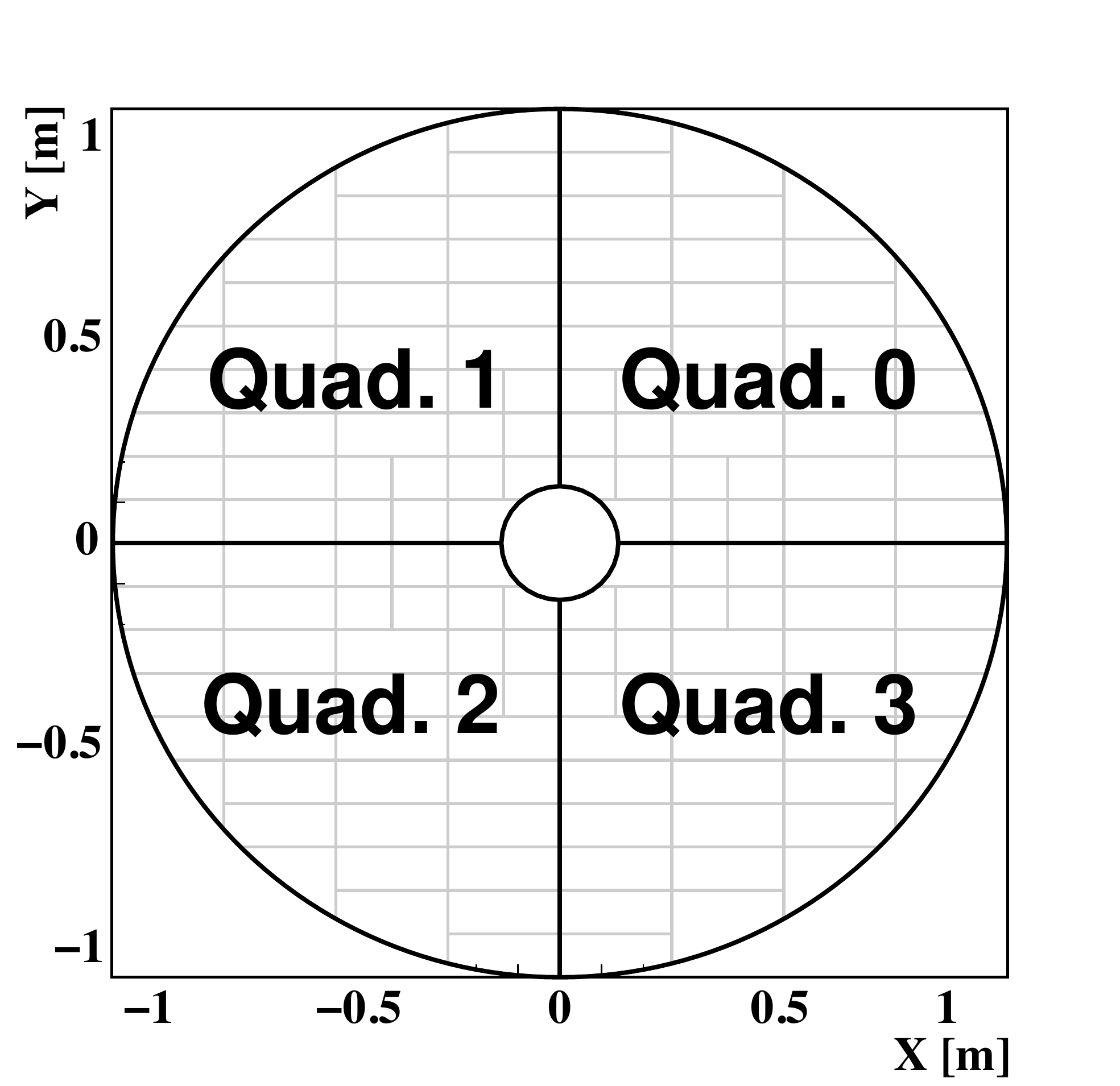}
\label{fig:quad:newchod}
\end{subfigure}
\begin{subfigure}[b]{0.45\textwidth}
\includegraphics[width=\textwidth]{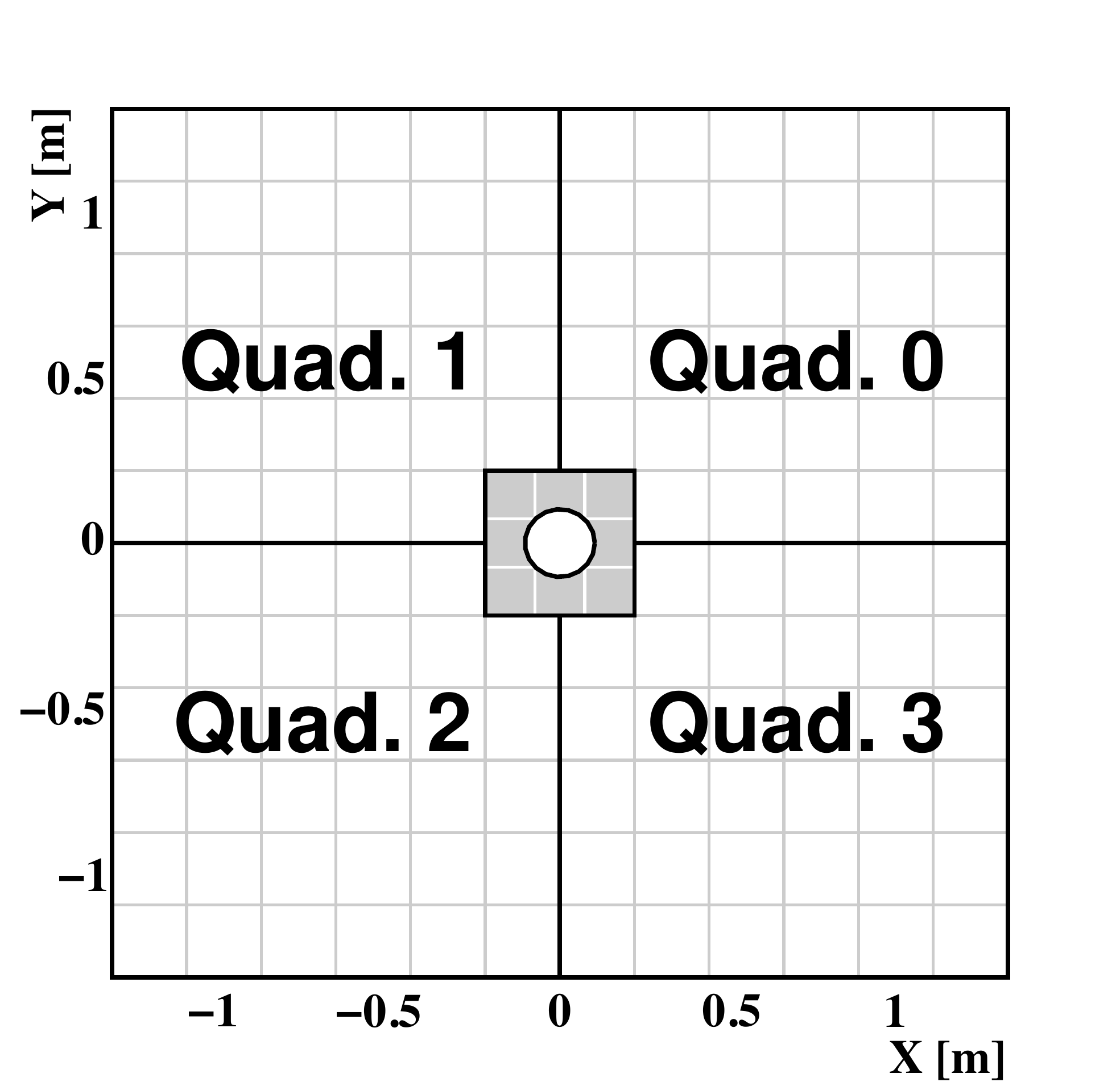}
\label{fig:quad:muv3}
\end{subfigure}
\caption{\label{fig:quad} Layout of the tiles and quadrants of the CHOD detector (left), and the tiles, quadrants, and inner/outer regions of the MUV3 detector (right). The smaller, inner MUV3 tiles are shown in grey; quadrants only involve the large \textit{outer} tiles shown in white. The hole in the centre of each detector allows the passage of the undecayed beam particles.}
\end{figure}

The hardware implementation of the L0 trigger system is described in detail in \cite{L0Paper}.
In total, five detectors are involved. 
The RICH, CHOD, NA48-CHOD and MUV3 participate via specialised firmware implemented in the readout hardware, with access to information from each readout channel. 
The CHOD, NA48-CHOD, and MUV3 define four regions as \textit{quadrants}, with the MUV3 excluding eight tiles close to the detector centre from the quadrant definition (Fig.~\ref{fig:quad}).
The LKr participates via a dedicated \textit{\lzCalo}\ system that uses information from groups of channels provided by calorimeter readout modules \cite{L0Calo2017}.

The data from each detector are processed to identify groups of coincident signals in the detector.
Each group is compared to a programmable set of up to 16 conditions per detector.
The RICH conditions are based on the total signal multiplicity, while the NA48-CHOD conditions are based on the number of quadrants in which there are signals.
The CHOD and MUV3 conditions are based on the total signal multiplicity as well as the number and pattern of quadrants in which there are signals.
The LKr conditions are based on the total energy deposited and the number of localised energy deposits (clusters).
The full list of L0 trigger conditions is given in Table~\ref{table:primID}.
If at least one of the conditions is satisfied in a given detector, a \emph{trigger primitive} is produced.
Each trigger primitive encodes which of the 16 conditions are satisfied, and a timestamp at \SI{100}{\pico\second} precision.

\begin{table}[ht]
\small
\centering
\renewcommand*\arraystretch{1.1}
\begin{tabular}{lll}
\hline
\textbf{Detector} & \textbf{Condition} & \textbf{Description} \\
\hline
RICH      & RICH      & At least two signals in the detector \\
\hline
CHOD      & Q1        & At least one signal in any quadrant\\
~         & Q2        & At least one signal in each of two different quadrants\\
~         & QX        & At least one signal in each of two diagonally-opposite quadrants\\
~         & UTMC      & Upper multiplicity condition: fewer than five signals in the detector\\
\hline
NA48-CHOD & NA48-CHOD & At least one signal in any quadrant \\
\hline
MUV3      & M1        & At least one signal in the detector\\
~         & MO1       & At least one signal in the outer tiles\\
~         & MO2       & At least two signals in the outer tiles\\
~         & MOQX      & At least one signal in each of two diagonally-opposite quadrants\\
\hline
LKr  & E10        & At least \SI{10}{\giga\eV} deposited in the LKr\\
~        & E20        & At least \SI{20}{\giga\eV} deposited in the LKr\\
~        & E30        & At least \SI{30}{\giga\eV} deposited in the LKr\\
~        & C2E5         & At least \SI{5}{\giga\eV} deposited in the LKr by at least two clusters\\
~        & LKr30      & Logical OR between E30 and C2E5\\
\hline
LKr  & E1        & At least \SI{1}{\giga\eV} deposited in the LKr\\
(beam dump)   & E2        & At least \SI{2}{\giga\eV} deposited in the LKr\\
~            & E4        & At least \SI{4}{\giga\eV} deposited in the LKr\\
~            & C2E2      & At least \SI{2}{\giga\eV} deposited in the LKr by at least two clusters\\
\hline
\end{tabular}
\caption{List of L0 trigger conditions. The LKr conditions differ between \kp\ and beam-dump configuration.}
\label{table:primID}
\end{table}

The trigger primitives are sent to the L0 trigger processor (L0TP)~\cite{L0TP}.
The L0TP produces triggers by combining primitives within \SI{6.25}{\nano\second} of those from a specified reference detector. The reference detector is either the RICH or NA48-CHOD in \kp\ mode, and either the CHOD or LKr in beam-dump configuration.
The time of the primitive from the reference detector defines the L0 trigger time.
The time resolutions of the primitives with respect to the RICH primitive are \SI{0.7}{\nano\second} for the CHOD and MUV3, and \SI{2}{\nano\second} for the NA48-CHOD and LKr.
These values are larger than those achieved offline, as precise time calibrations are not available at trigger level.

The combinations of primitives are compared to up to 16 criteria, which define the L0 part of each trigger line.
A trigger is generated if at least one of the criteria is satisfied.
Each trigger line can be downscaled individually, with a downscaling of $N$ indicating that one in $N$ generated triggers is transmitted.
The trigger lines and corresponding L0 trigger conditions for \kp\ mode are given in Table~\ref{table:L0physics}, and those for beam-dump configuration are given in Table~\ref{table:dumpmode}.

\begin{table}[h!]
\centering
\renewcommand*\arraystretch{1.2}
\begin{tabular}{lll}
\hline
\textbf{Trigger line} & \textbf{L0 trigger conditions} & \textbf{L1 trigger conditions}\\
\hline
PNN & \trig{RICH \cdot Q1 \cdot UTMC \,\cdot \overline{QX} \,\cdot\, \overline{M1} \,\cdot\, \overline{LKr30}} & \trig{KTAG \cdot \overline{LAV} \cdot STRAW} \\
Non-$\mu$ & \trig{RICH \cdot Q1 \,\cdot\, \overline{M1}} & \trig{KTAG \cdot \text{STRAW-1TRK}} \\ 
MT   & \trig{RICH \cdot QX} & \trig{KTAG \cdot \text{STRAW-Exo}} \\ 
2$\mu$MT       & \trig{RICH \cdot QX \cdot MO2} & \trig{KTAG \cdot \text{STRAW-Exo}} \\
$e$MT   & \trig{RICH \cdot QX \cdot E20} & \trig{KTAG \cdot \text{STRAW-Exo}} \\
$\mu$MT & \trig{RICH \cdot QX \cdot MO1 \cdot E10} & \trig{KTAG \cdot \overline{LAV} \cdot \text{STRAW-MT}} \\
DV-$\mu$      & \trig{RICH \cdot Q2 \cdot MO1 \cdot E10} & \trig{\overline{KTAG} \cdot \text{STRAW-DV}} \\
DV-$2\mu$  & \trig{RICH \cdot Q2 \cdot MO2 \,\cdot\, \overline{E10}} & \trig{\text{STRAW-DV}} \\
Neutrino         & \trig{RICH \cdot Q1 \cdot MOQX \,\cdot\, \overline{Q2}} & \trig{KTAG \cdot \overline{LAV} \cdot \text{STRAW-1TRK}}  \\
Control     & \trig{\text{NA48-CHOD}} & None\\
\hline
\end{tabular}
\caption{\kp\ trigger lines and the corresponding L0 and L1 trigger conditions. An overline indicates that the condition is used in veto.
}
\label{table:L0physics}
\end{table}

\begin{table}[h!]
\centering
\renewcommand*\arraystretch{1.2}
\begin{tabular}{ll}
\hline
\textbf{Trigger line} & \textbf{L0 trigger conditions} \\
\hline
Neutral & \trig{C2E2} \\ 
Charged & \trig{Q2} \\ 
Control & \trig{Q1} \\  
\hline
\end{tabular}
\caption{Beam-dump trigger lines and the corresponding L0 trigger conditions. No L1 conditions are applied in beam-dump configuration.}
\label{table:dumpmode}
\end{table}

\subsection{The high-level trigger}
\label{sec:L1}

The online framework consists of the DAQ software and the high-level trigger (HLT) software. 
The framework is deployed on the DAQ-farm, a cluster of 30 computers, with each computer independently executing both parts of the software on a subset of the L0-triggered events.

The data from all detectors except the GTK and calorimeters (LKr, MUV1, MUV2, IRC, SAC) are transmitted to the DAQ-farm following a positive L0 trigger decision.
The DAQ software collects and combines event fragments sent by the detectors to produce a L1 event ~\cite{Boretto2019TheCern}.
The HLT software executes the L1 trigger algorithms on the L1 event after applying detector-specific decoding and reconstruction.
If an event satisfies the L1 trigger requirements, the HLT software requests data from the GTK and calorimeters, which are then merged with the L1 event to produce a complete event with an average size of \SI{20}{\kilo\byte}.

The HLT software imposes sets of conditions from different algorithms in a specified order.
These conditions define the L1 part of each trigger line.
An HLT processor routine evaluates the trigger lines and executes the conditions in the defined order.
If an event does not fulfil a condition of a trigger line, the event is discarded without checking the following conditions.
A fraction (1\%) of the processed events are accepted by the HLT processor regardless of whether any trigger condition is satisfied. 
The resulting \emph{autopass} events are used to measure the L1 trigger efficiencies.

Three L1 algorithms -- KTAG, LAV and STRAW --
define the L1 trigger conditions.
Each algorithm is configurable, as discussed below, and the configuration can be different for each trigger line. 
The list of \kp\ trigger lines and corresponding L1 trigger conditions is given in Table~\ref{table:L0physics}. 
No L1 trigger conditions are applied in beam-dump configuration.

\subsubsection*{The L1 KTAG trigger}

The L1 KTAG trigger requires a \kp\ in time with the L0 trigger, which rejects events triggered at L0 by halo muons and scattered beam pions. The KTAG detects \cherenkov{} light produced by the \kp\ in 8 azimuthal sectors, each consisting of an array of 48 photomultipliers with a single-photon time resolution of \SI{280}{\pico\second}. 
As the average number of photons detected per \kp\ is 18, the KTAG measures the \kp\ time with a precision of \SI{70}{\pico\second} after offline corrections have been applied~\cite{KTAGPaper}.
The L1 KTAG trigger requires that more than four KTAG sectors have at least one detected \cherenkov{} photon within \SI{5}{\nano\second} of the L0 trigger time. 
The time window includes a safety margin to accommodate time drifts in the KTAG readout system.

\subsubsection*{The L1 LAV trigger}
The L1 LAV trigger identifies whether there is activity in the LAV detector in time with the L0 trigger, with the primary aim of vetoing \kpipi\ decays.
The LAV comprises 12 stations of ring-shaped electromagnetic calorimeters made of lead-glass blocks located inside and downstream of the vacuum tank, which detect photons with \SIlist[list-pair-separator = {--}, list-units=single]{1.0;1.5}{\nano\second} resolution.
The L1 LAV trigger requires at least three signals in stations 2--11 within \SI{6}{\nano\second} of the L0 trigger time.
Stations 1 and 12 are not considered to avoid vetoing events with inelastic interactions in the GTK and RICH detectors, respectively.

\subsubsection*{The L1 STRAW trigger}
The L1 STRAW trigger reconstructs the tracks of charged particles that traverse the STRAW detector in time with the L0 trigger.
The STRAW detector comprises four straw-chambers, two at either side of a dipole magnet, and measures the trajectories and momenta of charged particles. 
The L1 STRAW algorithm reconstructs tracks and performs a momentum computation by making~a two-dimensional Hough transform of signals in the four straw chambers~\cite{Hough}.
The relative momentum resolution of the L1 STRAW algorithm is $\Delta p/p \sim$ 1\%, about constant in the \SIlist[list-pair-separator = {--}, list-units=single]{15;45}{\giga\eV/\clight} momentum range of interest. This resolution is larger than 0.3\% achieved offline due to the simpler and faster reconstruction used at L1.
A preliminary sample of tracks is selected by requiring the track momentum to be at least \SI{3}{\giga\eV/\clight} and the track direction to satisfy the conditions 
$|dX/dZ|< 0.02$ and $|dY/dZ|< 0.02$.
To reduce the number of fake tracks, quality conditions are applied on the signal and track times.

The main L1 STRAW trigger condition is optimised for \kpinunu\ decays. It aims to reject multi-track events and one-track events with tracks originating outside the \FV\ or with a momentum outside the range used for \kpinunu\ event selection.
The track selection requires: 
positive charge; 
momentum below \SI{50}{\giga\eV/\clight};
closest distance of approach (CDA) of the track to the nominal beam axis below \SI{200}{\milli\metre};
and Z position at the CDA upstream of the first straw chamber.
Multi-track events are identified by the presence of a pair of tracks with CDA between the two tracks below \SI{30}{\milli\meter}, which is consistent with the two tracks originating from the same \kp\ decay. 
An event is accepted if it contains at least one track satisfying the track selection, and is not identified as a multi-track event.

Several other L1 STRAW conditions share the same pattern recognition and track reconstruction, only differing in the requirements on the reconstructed tracks: 
L1 STRAW-1TRK is a less restrictive version of the L1 STRAW condition that selects final states with at least one positively-charged track originating in the \FV\ with momentum below \SI{65}{\giga\eV/\clight}; 
multi-track final states are collected by L1 STRAW-Exo, which requires at least one negatively-charged track, and L1 STRAW-MT, which identifies events with at least three tracks originating in the \FV; 
and L1 STRAW-DV collects pairs of tracks originating from a displaced vertex (DV) with CDA position at least \SI{100}{\milli\meter} from the nominal beam axis, indicating the decay of a secondary long-lived particle.

\section{The trigger performance}
\label{sec:performance}

Throughout 2016--2018, fluctuations in the instantaneous beam intensity during the spill caused
the L0 trigger rate to rise above the \SI{1}{\mega\hertz} limit of the L0TP.
To avoid this, the beam intensity was set at 60\% of the nominal value, corresponding to a mean instantaneous intensity of \SI{450}{\mega\hertz}, or $1.2\times 10^{9}$ Argonion counts.
Late in 2018, a \emph{choke} system was developed that enables each readout system and the L0TP to temporarily suppress L0 triggers if the trigger rate is too high.
The choke system allowed operation at 80\% of the nominal beam intensity towards the end of 2018.

\subsection{Trigger rates}

\begin{figure}[t!]
\centering
\begin{subfigure}[b]{0.48\textwidth}
\includegraphics[width=\textwidth]{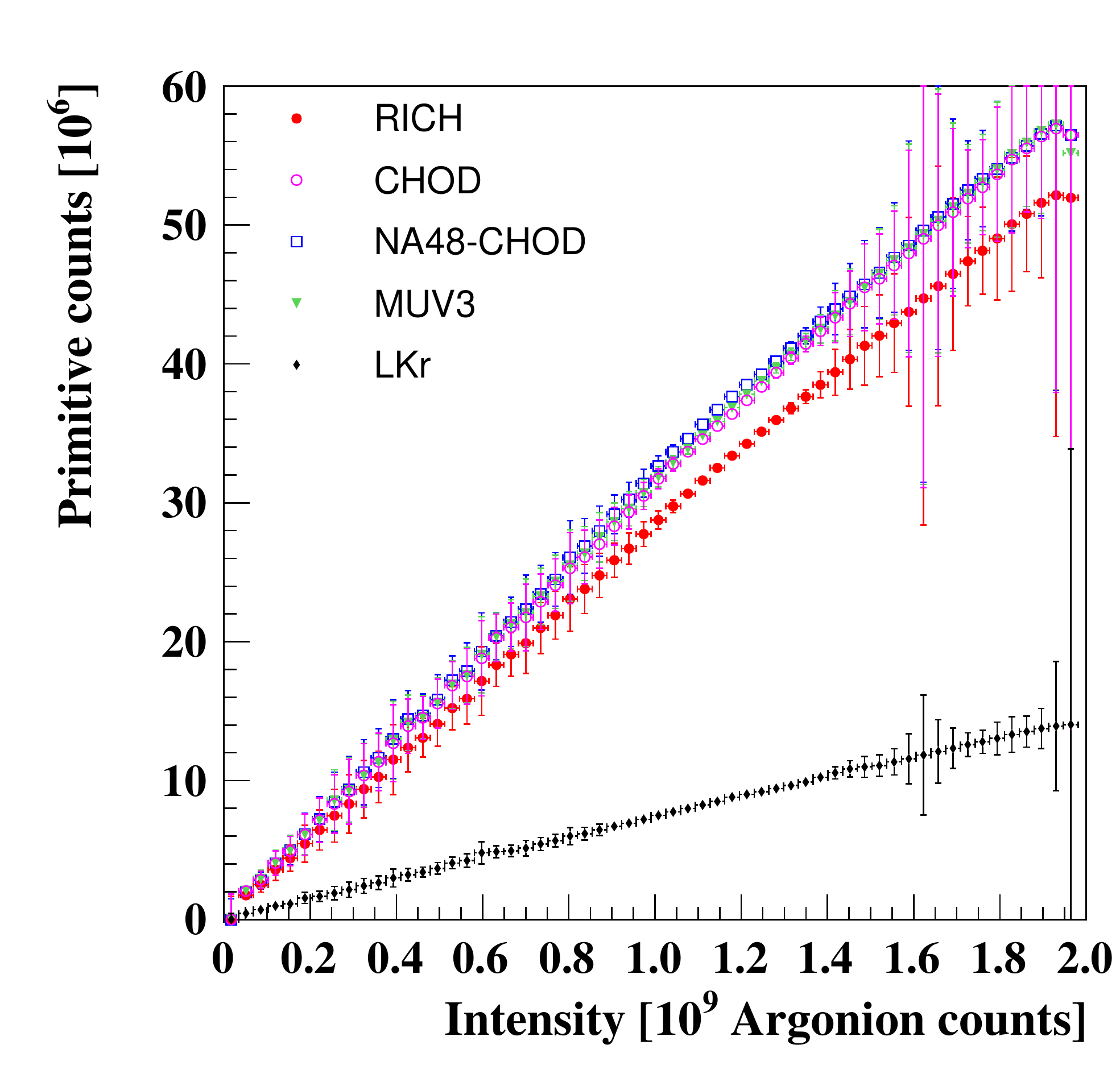}
\end{subfigure}
\hfill
\begin{subfigure}[b]{0.48\textwidth}
\includegraphics[width=\textwidth]{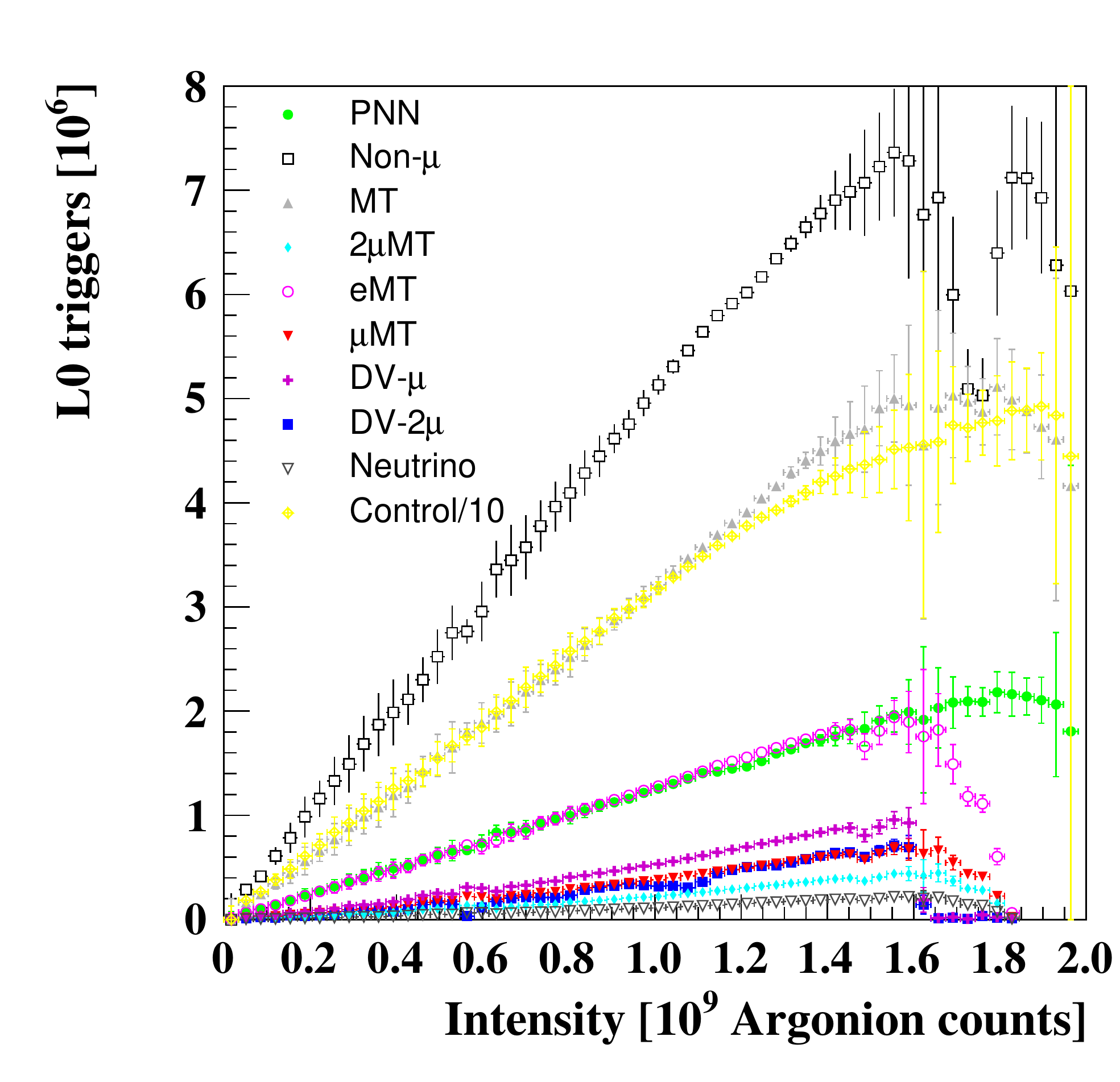}
\end{subfigure}
\caption{Number of primitives (left) and L0 triggers before downscaling (right) per spill as a function of the average beam intensity. Local variations in the data distributions are within statistical uncertainties.}
\label{fig:RatesVsIntensity}
\end{figure}

The number of L0 primitives produced by each detector is expected to rise linearly with the average beam intensity. 
The number of L0 primitives received by the L0TP is shown as a function of the average beam intensity in the left panel of Fig.~\ref{fig:RatesVsIntensity}.
At \num{1.2e9} Argonion counts, the CHOD (NA48-CHOD) produces \num{3.7e7} (\num{3.8e7}) primitives per spill.
The MUV3, despite only being traversed by muons, also produces \num{3.8e7} primitives per spill; this is explained by additional muons from $\pip \to \mup \nu$ decays of pions in the beam, which do not enter the other detectors.
The RICH is insensitive to particles below the \cherenkov\ threshold, and produces \num{3.4e7} primitives per spill.
The LKr is insensitive to minimum ionising particles, and produces \num{9e6} primitives per spill.

The number of L0 triggers generated by each L0 trigger line per spill is shown as a function of the average beam intensity in the right panel of Fig.~\ref{fig:RatesVsIntensity}.
The number of generated triggers increases linearly with average beam intensity below $1.2 \times 10^{9}$ Argonion counts, and saturates above this value.
The saturation is caused by the rate of primitives temporarily exceeding the maximum \SI{17.3}{\mega\hertz} input rate of the L0TP due to fluctuations in the instantaneous beam intensity; this has no significant effect on the data acquisition.
For each trigger line, the downscaling factor and typical number of triggers per spill after downscaling are shown in Table~\ref{table:L1physics}.
The total number of L0 triggers per spill is \num{2.2e6}, lower than the sum of the entries in Table~\ref{table:L1physics} as the same event can be triggered by more than one line.
The number of L1 triggers generated by each trigger line per spill is shown in Table~\ref{table:L1physics}.
The total number of L1 triggers per spill is $2.5\times10^{5}$, with the majority of events collected by the PNN and control trigger lines. 

\begin{table}[h!]
\centering
\renewcommand*\arraystretch{1.2}
\begin{tabular}{lccc}
\textbf{Trigger line} & \textbf{Downscaling} & \textbf{L0 triggers [$10^{3}$]} & \textbf{L1 triggers [$10^{3}$]}\\
\hline
PNN & 1 & 1540 & 74\\
Non-$\mu$ & 200 & 30 & 12\\
MT & 100 & 39 & 4\\   
2$\mu$MT & 2 & 150 & 30\\
$e$MT & 8 & 193 & 22\\
$\mu$MT & 5 & 99 & 10\\
DV-$\mu$ & 5 & 140 & 0.3\\
DV-$2\mu$ & 3 & 160 & 5\\
Neutrino & 15 & 10 & 3\\
Control & 400 & 94 & 94\\
\end{tabular}
\caption{List of \kp\ trigger lines, L0 downscaling factors, and numbers of L0 triggers after downscaling and L1 triggers produced per spill at a beam intensity of \num{1.2e9} Argonion counts. No L1 conditions are applied to the control trigger line. }
\label{table:L1physics}
\end{table}

\begin{table}[h!]
\centering
\renewcommand*\arraystretch{1.2}
\begin{tabular}{lcccc}
\textbf{Trigger line} & \textbf{Downscaling} & \textbf{L0 triggers [$10^{3}$]} \\
\hline
Charged & 1 & 1\\
Neutral & 1 & 30\\
Control & 5 & 180\\
\end{tabular}
\caption{List of beam-dump trigger lines, L0 downscaling factors, and numbers of L0 triggers produced per spill at a beam rate of $10^{12}$ POT/s.
}
\label{table:beamdump2}
\end{table}

In beam-dump configuration only the L0 trigger is utilised.
The downscaling factors and numbers of L0 triggers per spill are shown in Table~\ref{table:beamdump2}. 
About $2.1\times10^{5}$ events are collected per spill, with the majority collected by the control trigger line.

\subsection{Efficiency of the PNN trigger line}
\label{subsec:l0pnn}
\begin{figure}[t]
\centering
\begin{subfigure}[b]{0.48\textwidth}
\includegraphics[width=\textwidth]{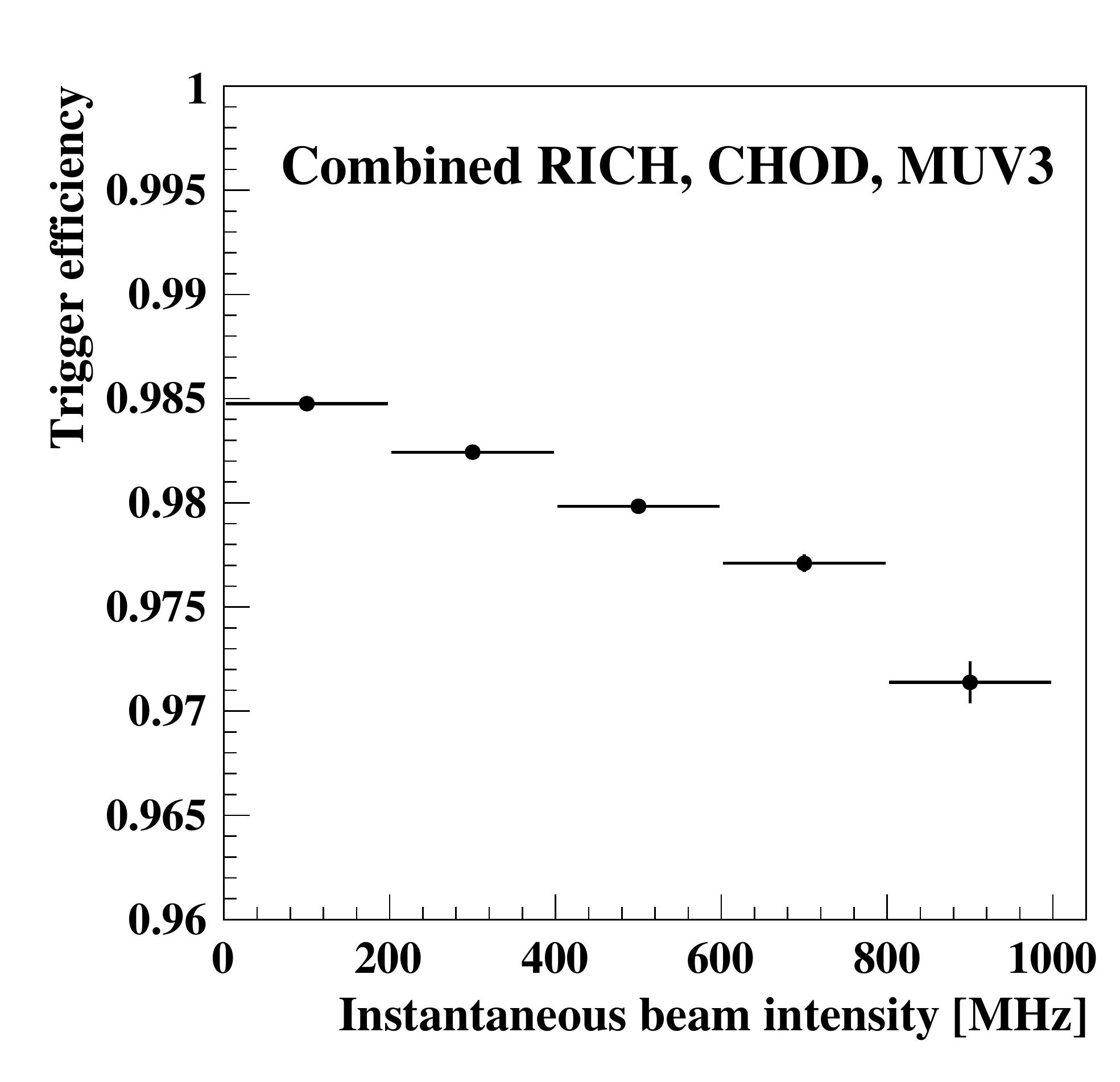}
\end{subfigure}
\begin{subfigure}[b]{0.48\textwidth}
\includegraphics[width=\textwidth]{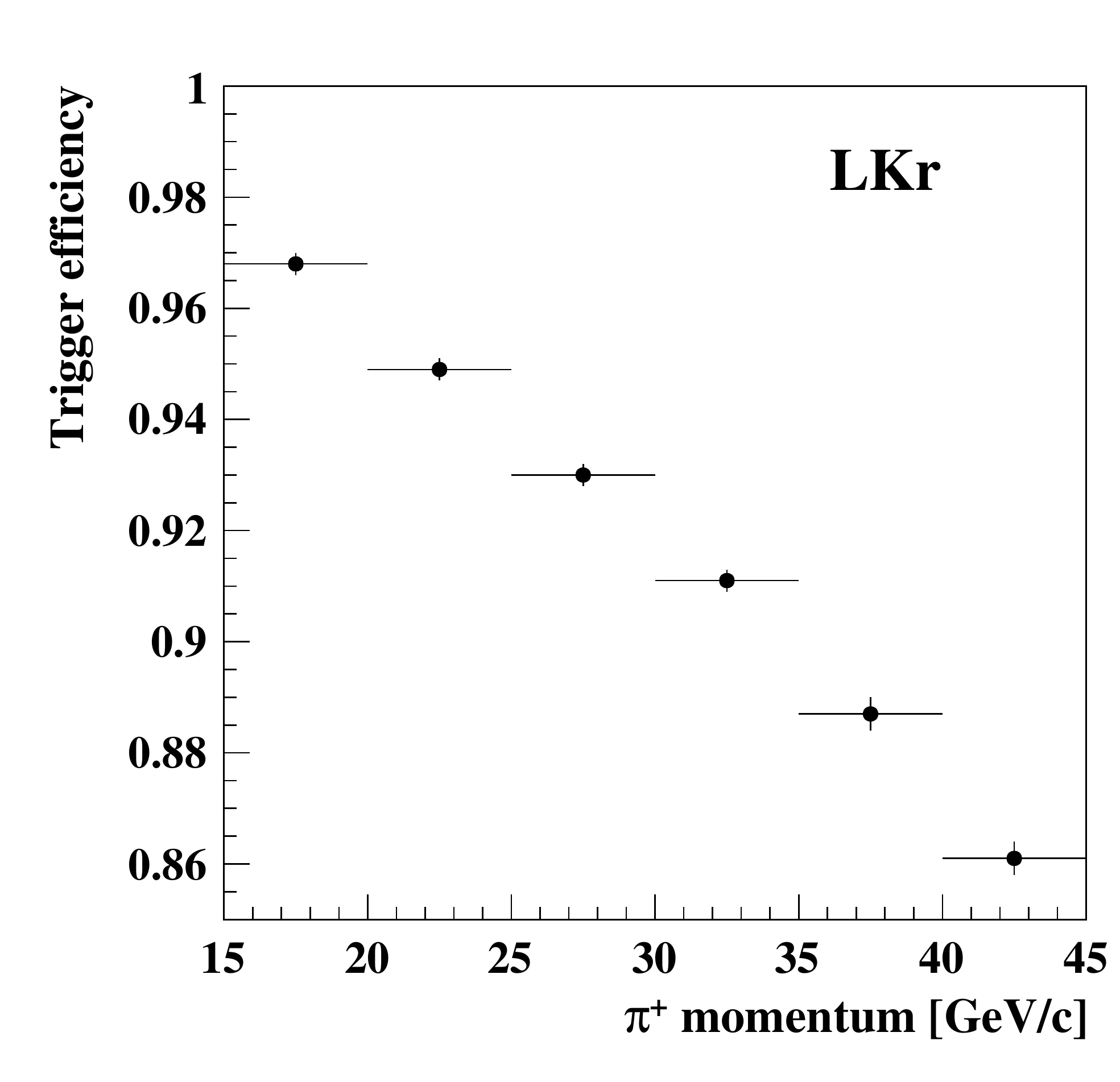}
\end{subfigure}
\caption{Efficiency of the combined RICH, CHOD, and MUV3 conditions of the PNN trigger line evaluated using \kpipi\ events as a function of instantaneous beam intensity (left), and efficiency of the LKr condition of the PNN trigger line evaluated using \kpipi\ events as a function of the \pip\ momentum (right).}
\label{fig:L0PNNTrigger}
\end{figure}

\begin{figure}[ht]
\centering
\includegraphics[width=\textwidth]{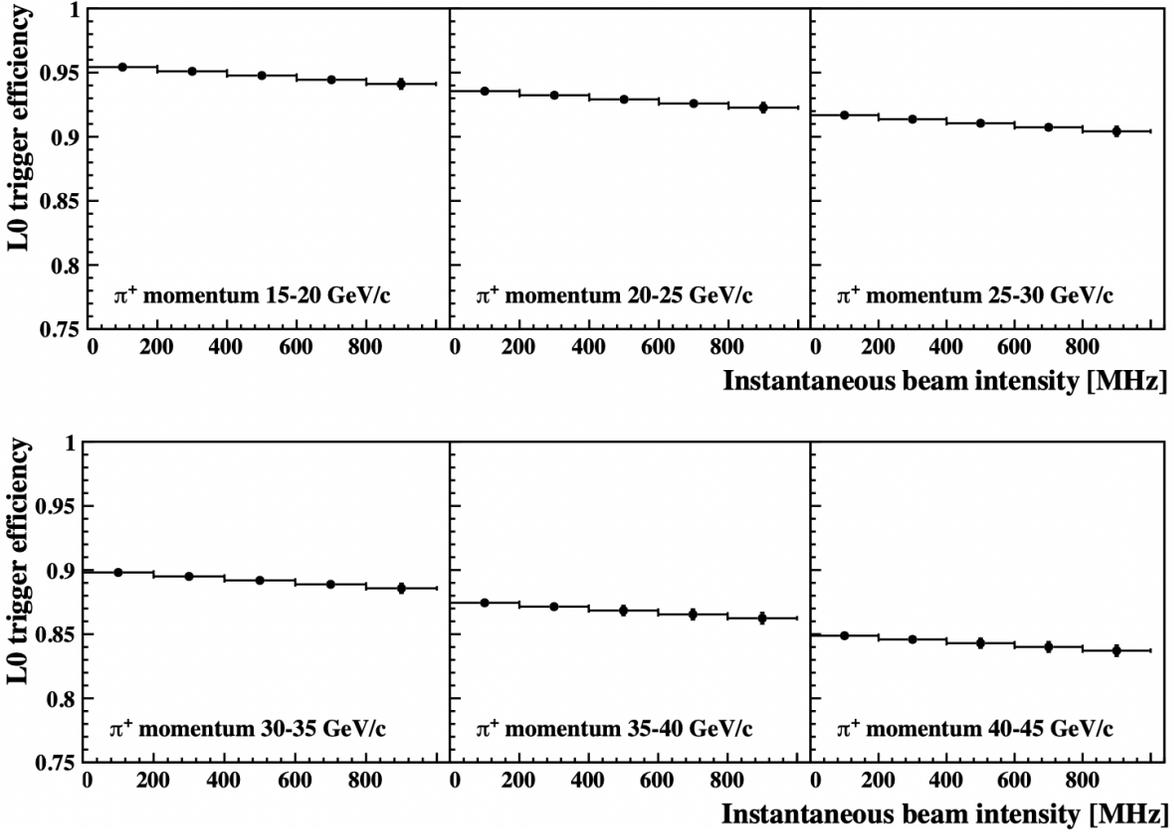}
\caption{\label{fig:L0PnnTriggerTot} Efficiency of the PNN L0 trigger conditions as a function of the instantaneous beam intensity in bins of \pip\ momentum. }
\end{figure}

In the context of the \kpinunu\ analysis, 
the combined efficiency of the L0 trigger conditions for RICH, CHOD, and MUV3 (Table~\ref{table:L0physics}) is measured using a sample of \kpipi\ events selected from minimum-bias data using criteria based on the \kpinunu\ event selection~\cite{Pnn2017}.
The efficiency is 98\% at the mean instantaneous beam intensity of \SI{450}{\mega\hertz},
 and varies with instantaneous beam intensity (left panel of Fig.~\ref{fig:L0PNNTrigger}).
The inefficiency, which arises because the time window considered at L0 is larger than the offline one, is mainly due to accidental activity in the MUV3 detector used in a veto condition.
The efficiency of the L0 trigger condition of the LKr is measured with a sample of \kpipi\ decays with the two photons from the \pgg\ detected in the LAV, using \pip\ from $\kpipipi$ decays to determine the relationship between \pip\ momentum and the energy deposited in the LKr~\cite{Pnn2017}.
The efficiency decreases from 96.5\% to 86\% over the \SIlist[list-pair-separator = {--}, list-units=single]{15;45}{\giga\eV/\clight} \pip\ momentum range relevant to the \kpinunu\ analysis
(right panel of Fig.~\ref{fig:L0PNNTrigger}).
The main part of the inefficiency stems from the requirement that the total energy deposited in the LKr is less than \SI{30}{\giga\eV}.
The overall L0 trigger efficiency is computed as the product of the two independent efficiencies discussed above. 
The~efficiency~varies~with~instantaneous~beam~intensity~and~\pip~momentum~(Fig.~\ref{fig:L0PnnTriggerTot}). 

\begin{figure}[h!]
\centering
\begin{subfigure}[h]{0.48\textwidth}
\includegraphics[width=\textwidth]{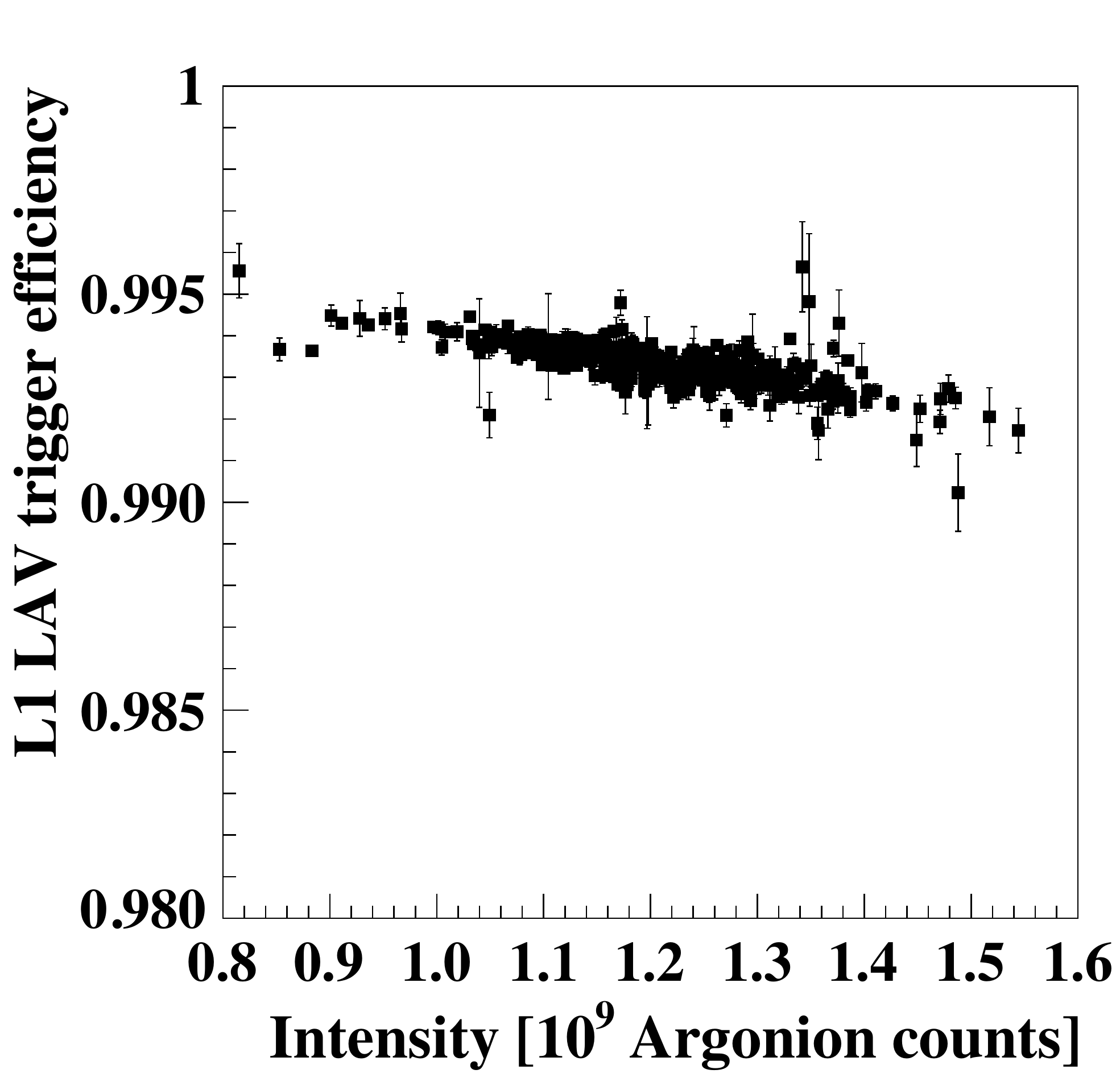}
\end{subfigure}
\begin{subfigure}[h]{0.48\textwidth}
\includegraphics[width=\textwidth]{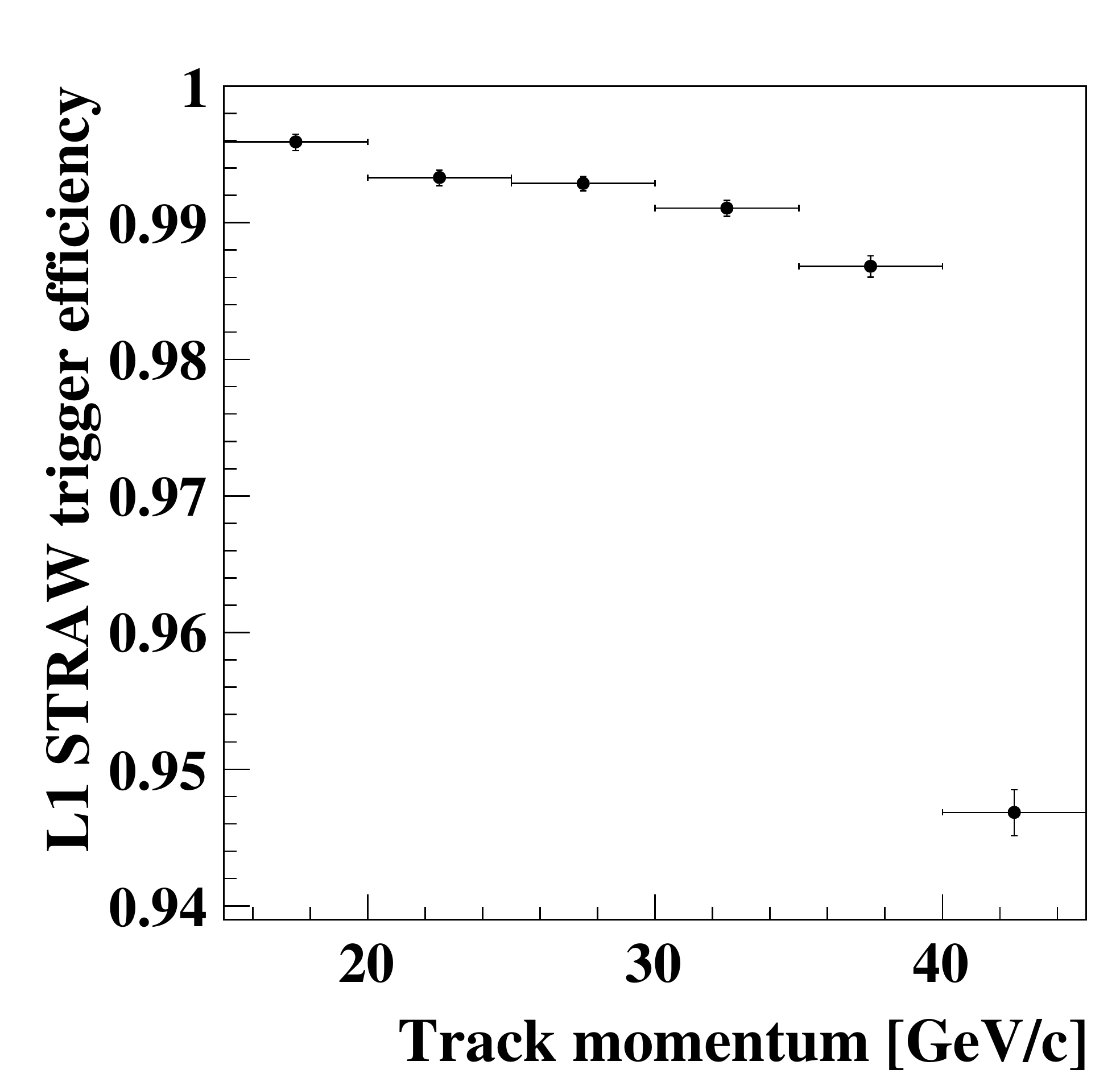}
\end{subfigure}
\caption{Efficiency of the L1 LAV trigger evaluated using \kmunu\ events as a function of the average beam intensity (left), and efficiency of the L1 STRAW trigger condition evaluated using \kmunu\ events as a function of the track momentum (right).}
\label{fig:L1PNNTrigger}
\end{figure}

The efficiencies of the L1 KTAG, LAV, and STRAW trigger conditions are measured using a sample of $\kmunu$ events selected from minimum-bias data using the same criteria for the KTAG, LAV, and STRAW as in the \kpinunu\ selection. 
The L1 KTAG efficiency is 99.9\%.
The L1 LAV efficiency ranges from 99.5\% to 99.0\% depending on the average beam intensity (left panel of Fig.~\ref{fig:L1PNNTrigger}); 
the inefficiency is due to random veto, as the time window used at L1 is larger than in the event selection.  
The efficiency of L1 STRAW varies between 99.6\% and 94.6\% as a function of the track momentum (right panel of Fig. \ref{fig:L1PNNTrigger}).
Non-gaussian tails in the online momentum resolution reduce the efficiency in the 40--\SI{45}{\giga\eV/\clight} region.
The total L1 trigger efficiency is computed as the product of the individual L1 KTAG, LAV, and STRAW efficiencies, and is 98\% (94\%) for \pip\ momentum in the 15--\SI{40}{\giga\eV/\clight} (40--\SI{45}{\giga\eV/\clight}) range.

The overall trigger efficiency 
is computed as the product of the L0 and L1 trigger efficiencies 
and is found to be 90\%, varying by 5\% both as a function of instantaneous beam intensity and \pip\ momentum.

\subsection{Efficiencies of the other \kp\ trigger lines}
\label{L0MultiSection}

\begin{figure}[t!]
\centering
\begin{subfigure}[b]{0.48\textwidth}
\includegraphics[width=\textwidth]{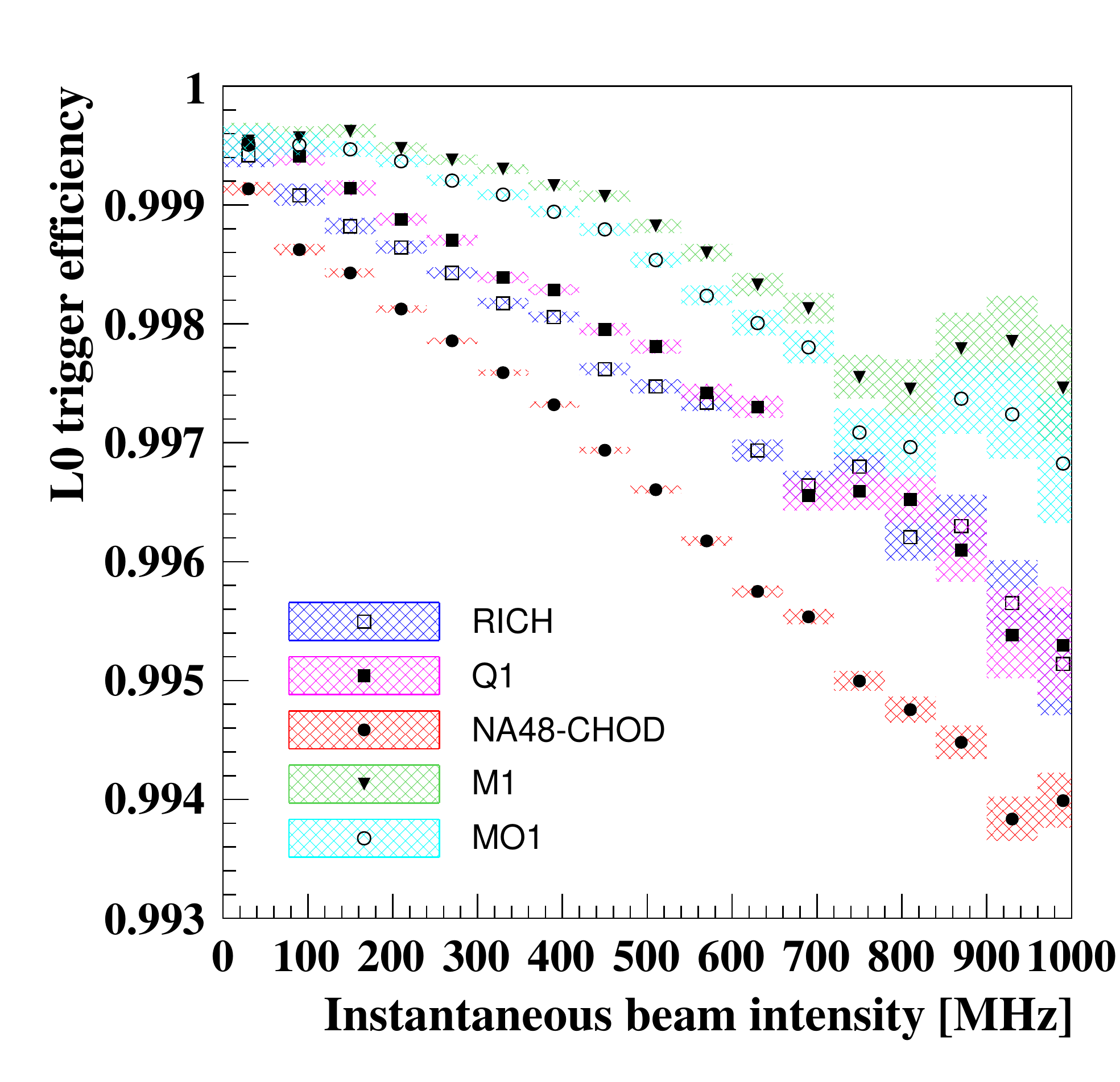}
\label{fig:K3pi-L0-Single}
\end{subfigure}
\begin{subfigure}[b]{0.48\textwidth}
\includegraphics[width=\textwidth]{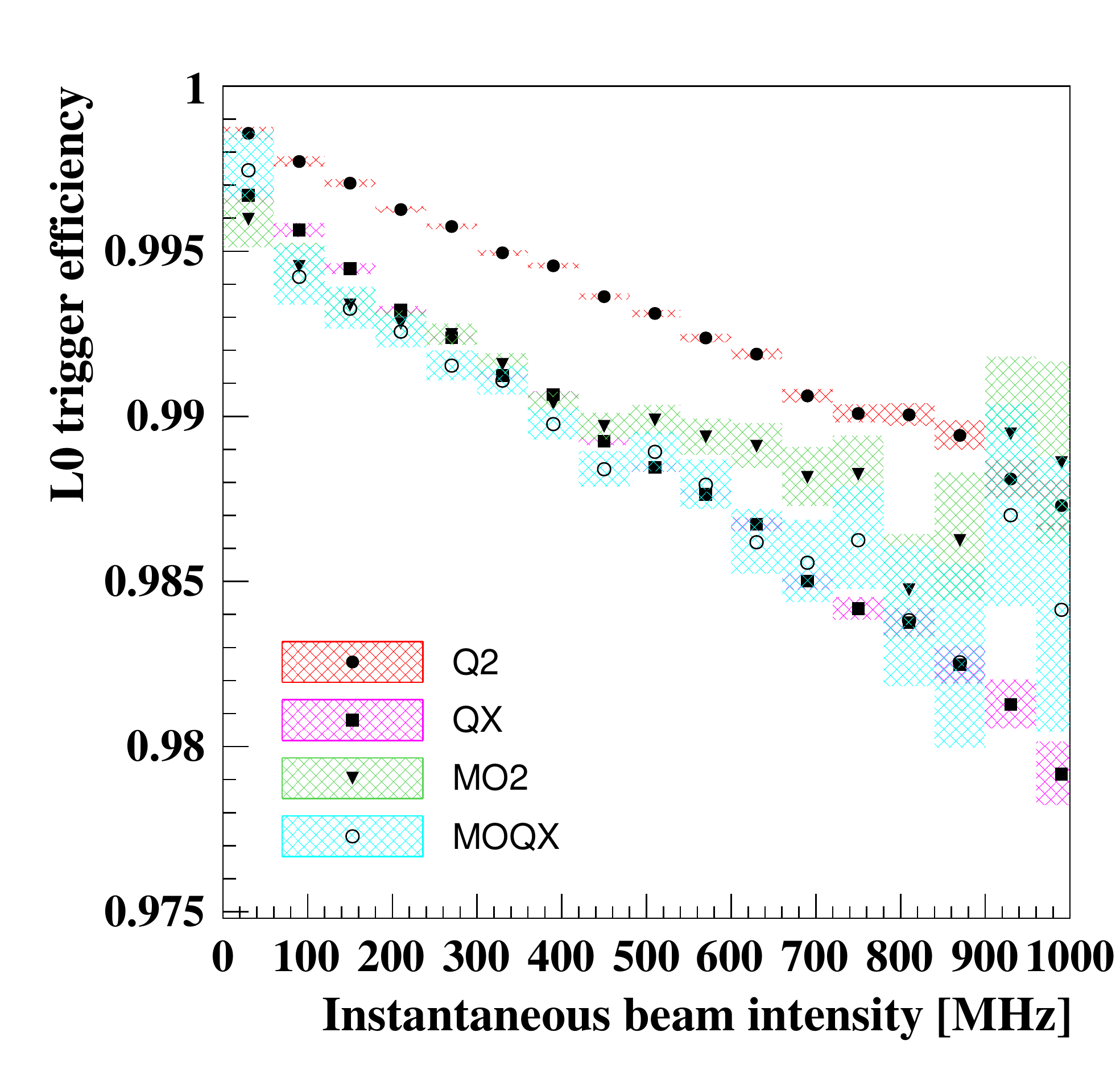}
\label{fig:K3pi-L0-Multi}
\end{subfigure}
\caption{Efficiencies of the L0 trigger conditions measured as a function of the instantaneous beam intensity using a sample of \kpipipi\ decays. The hatched areas indicate the statistical uncertainties of the measurements.}
\label{fig:K3pi-L0}
\end{figure}

The efficiencies of the L0 and L1 trigger conditions used in the other \kp\ trigger lines (Table~\ref{table:L0physics}) are measured using a sample of \kpipipi\ decays collected via the MT and control trigger lines.
This event sample also allows the study of trigger conditions involving muons from $\pi^\pm~\to~\mu^\pm~\nu$\ decays upstream of the LKr.
The efficiencies of a subset of the L0 conditions are shown as a function of the instantaneous beam intensity in Fig.~\ref{fig:K3pi-L0}.
At the mean instantaneous beam intensity of \SI{450}{\mega\hertz}, the RICH, NA48-CHOD, Q1, M1, and MO1 conditions have efficiencies above 99.7\%,
while the Q2, QX, MO2 and MOQX conditions have efficiencies above 99\%.
The efficiencies all decrease with increasing instantaneous beam intensity.

\begin{figure}[ht]
\centering
\includegraphics[width=0.5\textwidth]{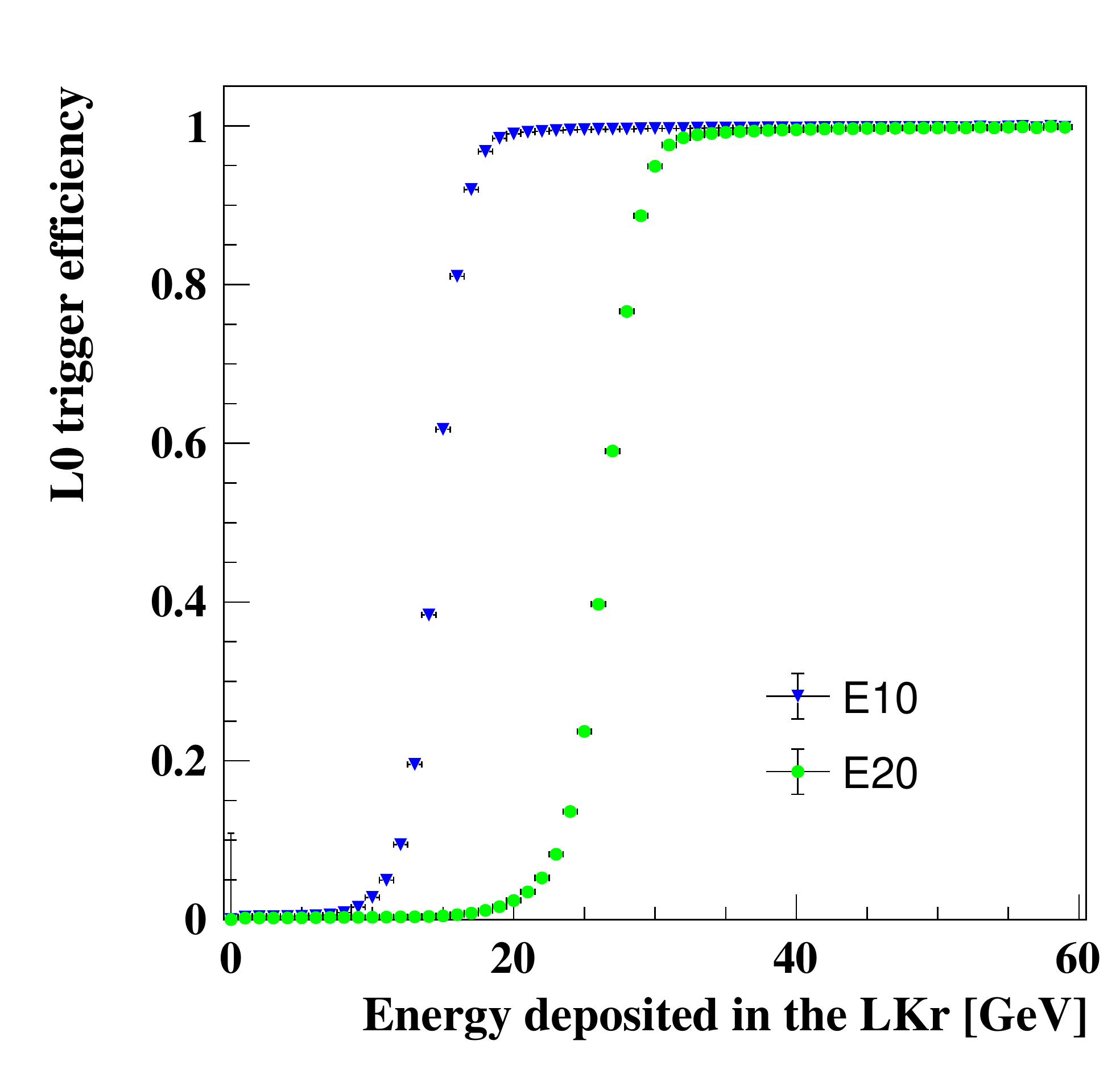}
\label{fig:K3pi-L0-LKr}
\caption{
Efficiencies of the E10 and E20 trigger conditions as a function of LKr energy deposit, measured with a sample of \kpipipi\ decays.}
\label{fig:K3pi-L0-UTMC-LKr}
\end{figure}

The efficiencies of the LKr conditions E10 and E20 are measured using \kpipipi\ events in which each reconstructed cluster is associated with one of the tracks and is within \SI{6}{\nano\second} of the L0 trigger time.
This requirement ensures that the energy reconstructed offline is assigned to a single LKr primitive.
Both conditions have efficiencies above 99.5\% in the plateau region of the turn-on curve (Fig.~\ref{fig:K3pi-L0-UTMC-LKr}).

\begin{figure}[h!]
\centering
\begin{subfigure}[b]{0.48\linewidth}
\includegraphics[width=\textwidth]{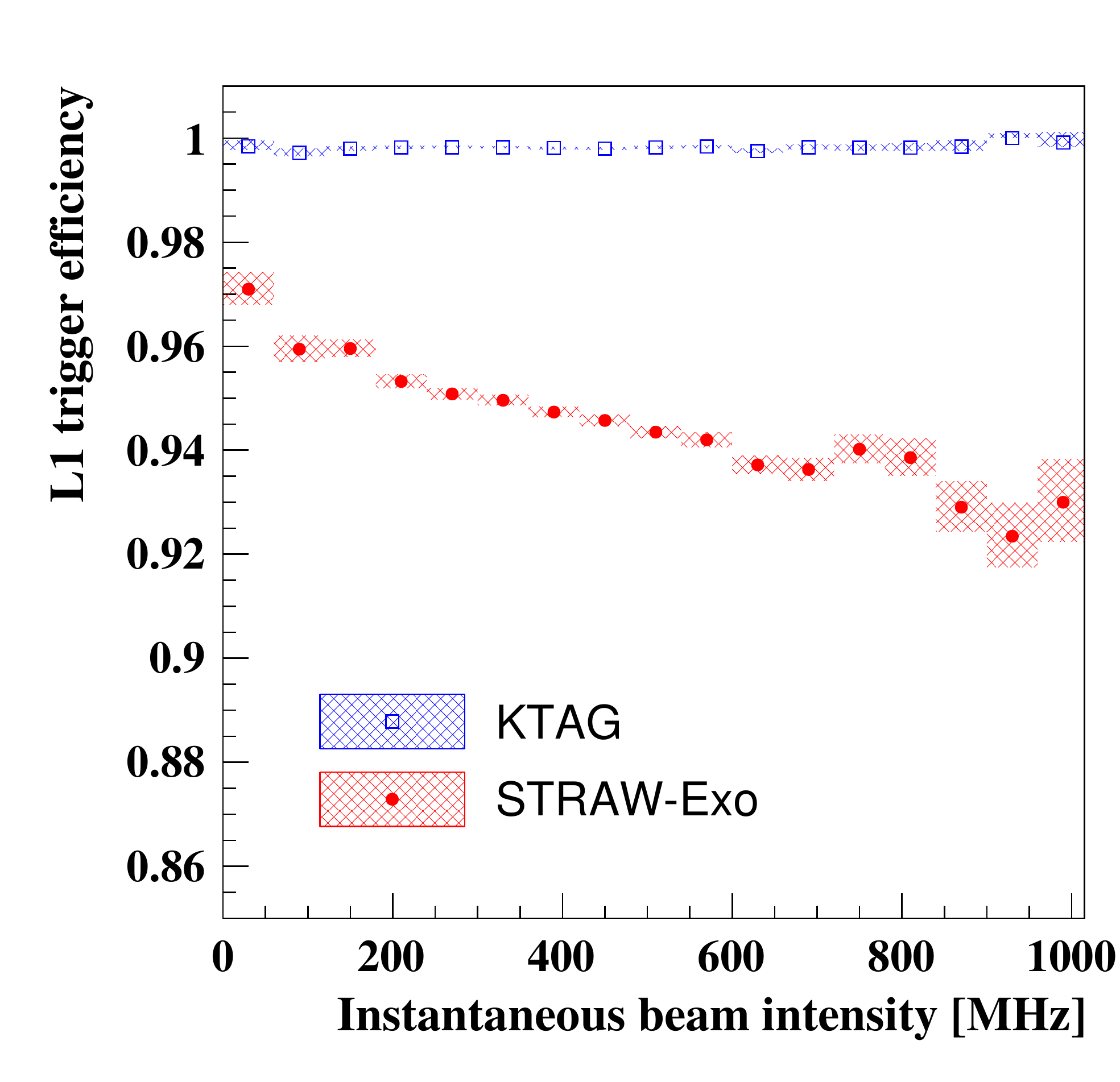}
\label{fig:K3pi-L1-KTAGSTRAWExo}
\end{subfigure}
\hfill
\begin{subfigure}[b]{0.48\textwidth}
\includegraphics[width=\textwidth]{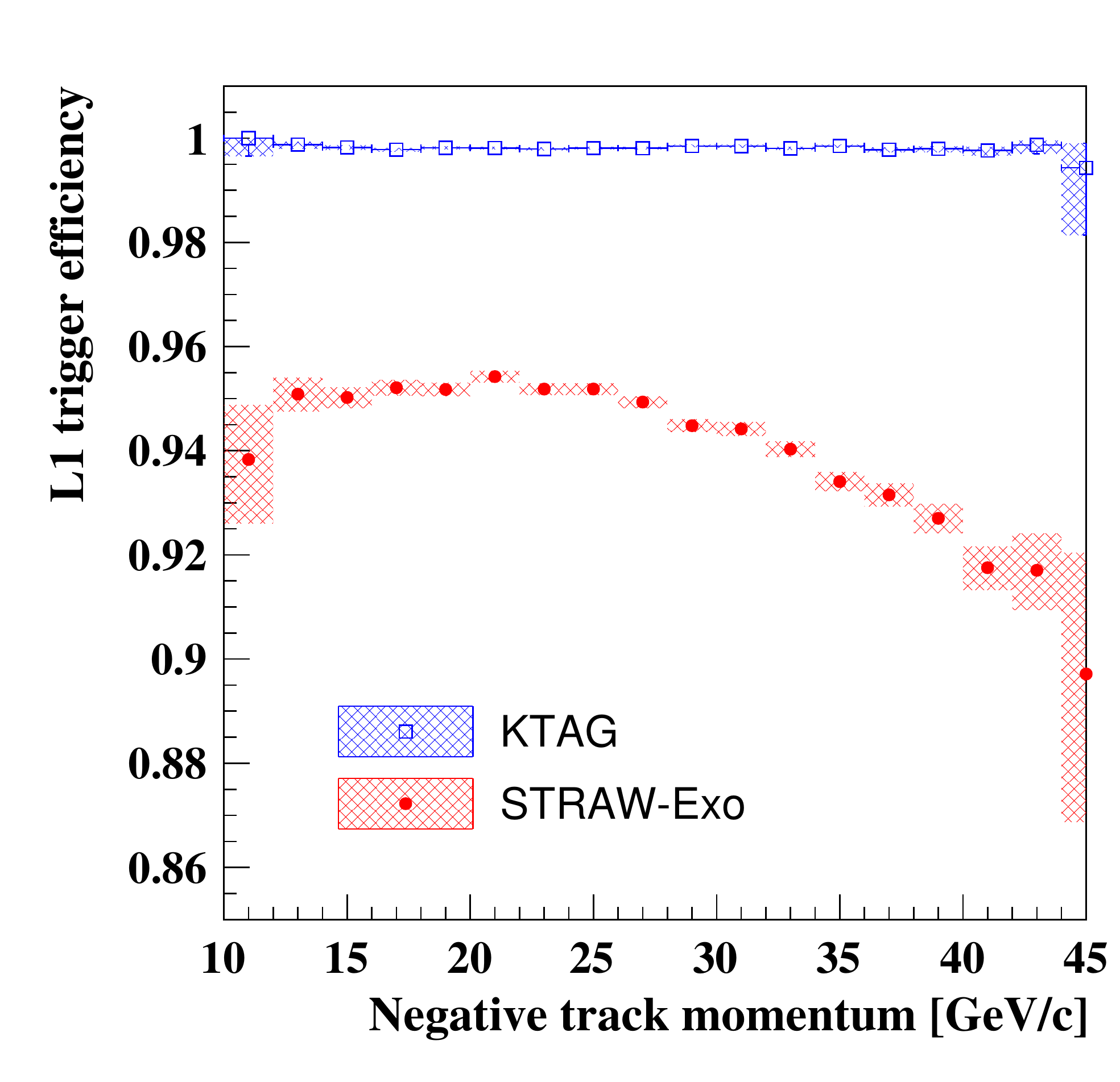}
\label{}
\end{subfigure}
\caption{L1 KTAG and STRAW-Exo efficiencies measured with a sample of \kpipipi\ decays as functions of instantaneous beam intensity (left) and \pim\ momentum (right). The hatched areas indicate the statistical uncertainties of the measurements.}
\label{fig:K3pi-L1}
\end{figure}

The efficiencies of the L1 KTAG and STRAW-Exo conditions are evaluated using autopass events as a function of the instantaneous beam intensity and the \pim\ momentum
(Fig.~\ref{fig:K3pi-L1}). 
The efficiency of L1 KTAG is 99.9\%, with no dependence on either variable, while L1 STRAW-Exo is 95\% efficient with a dependence on both.
The inefficiency of L1 STRAW-Exo is due to differences in the track reconstructions used at L1 and offline.

\subsection{Efficiencies of the beam-dump trigger lines}\label{sec:beamdump}

\begin{figure}[t]
\centering
\begin{subfigure}[h]{0.48\textwidth}
\includegraphics[width=\textwidth]{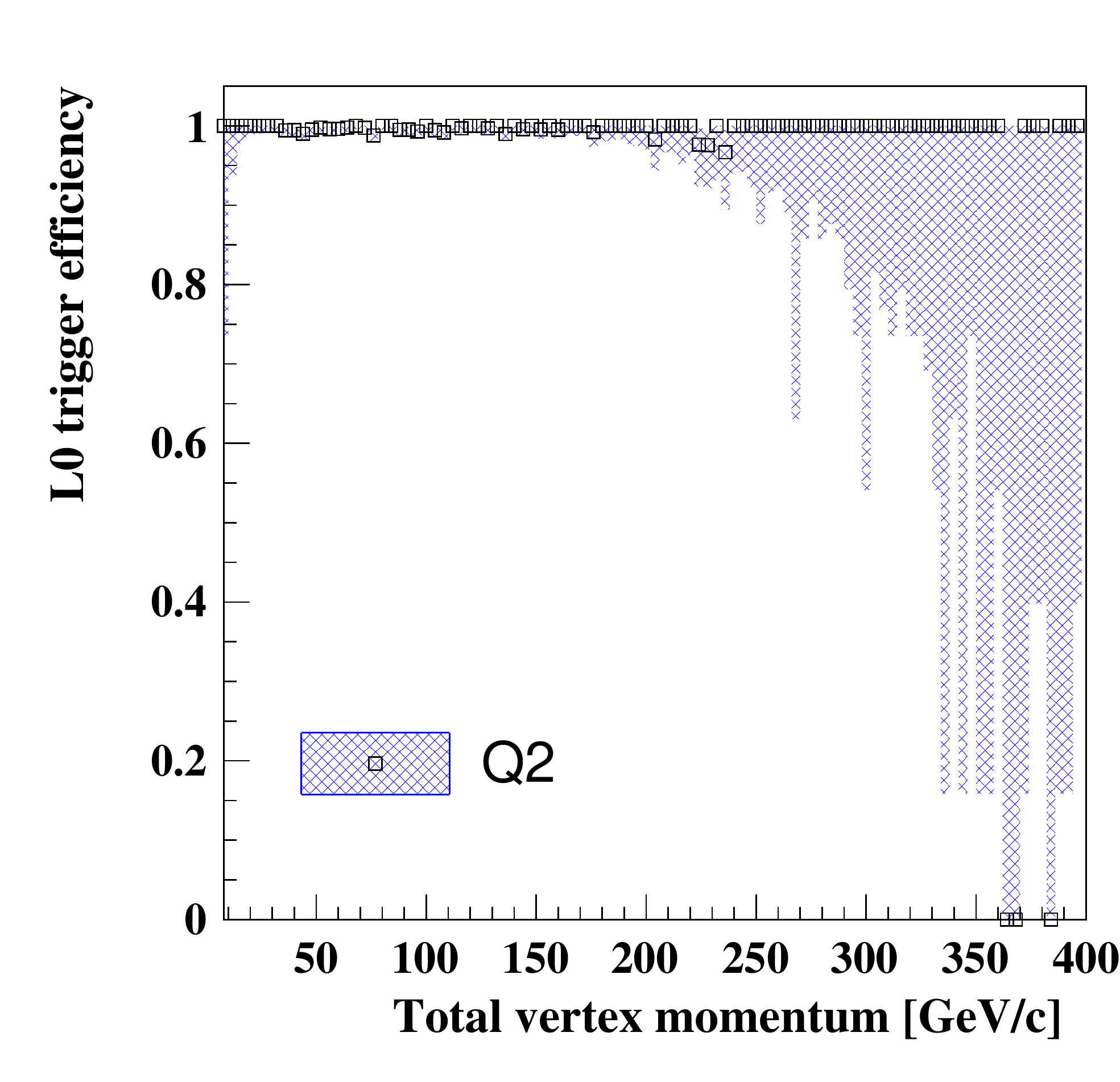}
\label{fig:newchodprim:left}
\end{subfigure}
\begin{subfigure}[h]{0.48\textwidth}
\includegraphics[width=\textwidth]{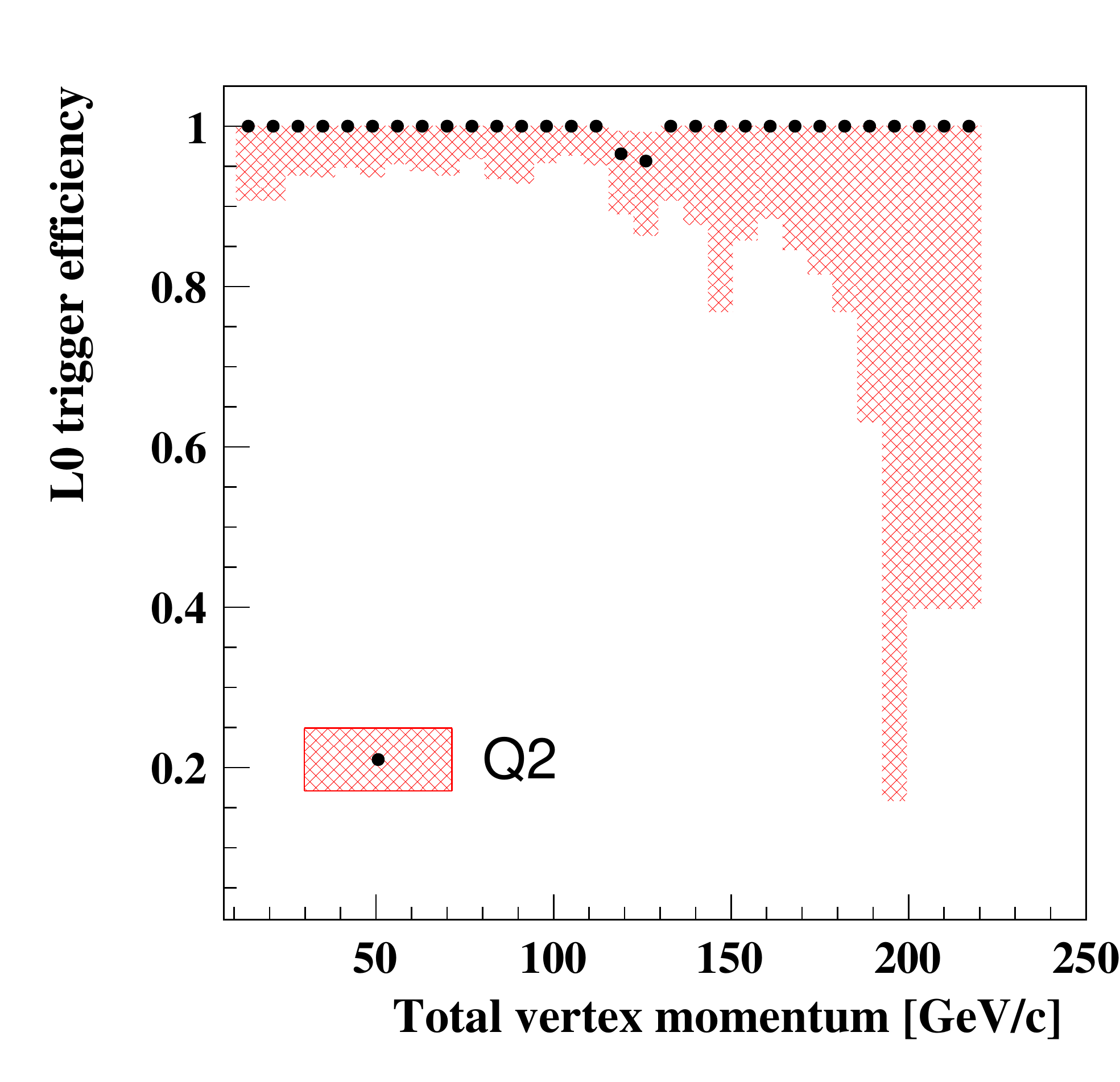}
\label{fig:newchodprim:right}
\end{subfigure}
\caption{Efficiency of the Q2 trigger condition in beam-dump data, measured as a function of the total vertex momentum using events collected via the control trigger (left) or the neutral trigger (right) lines reported in Table.~\ref{table:dumpmode}. The hatched areas indicate the statistical uncertainties of the measurements.}
 \label{fig:newchodprim}
\end{figure}

\begin{figure}[h!]
\centering
\includegraphics[width=0.5\textwidth]{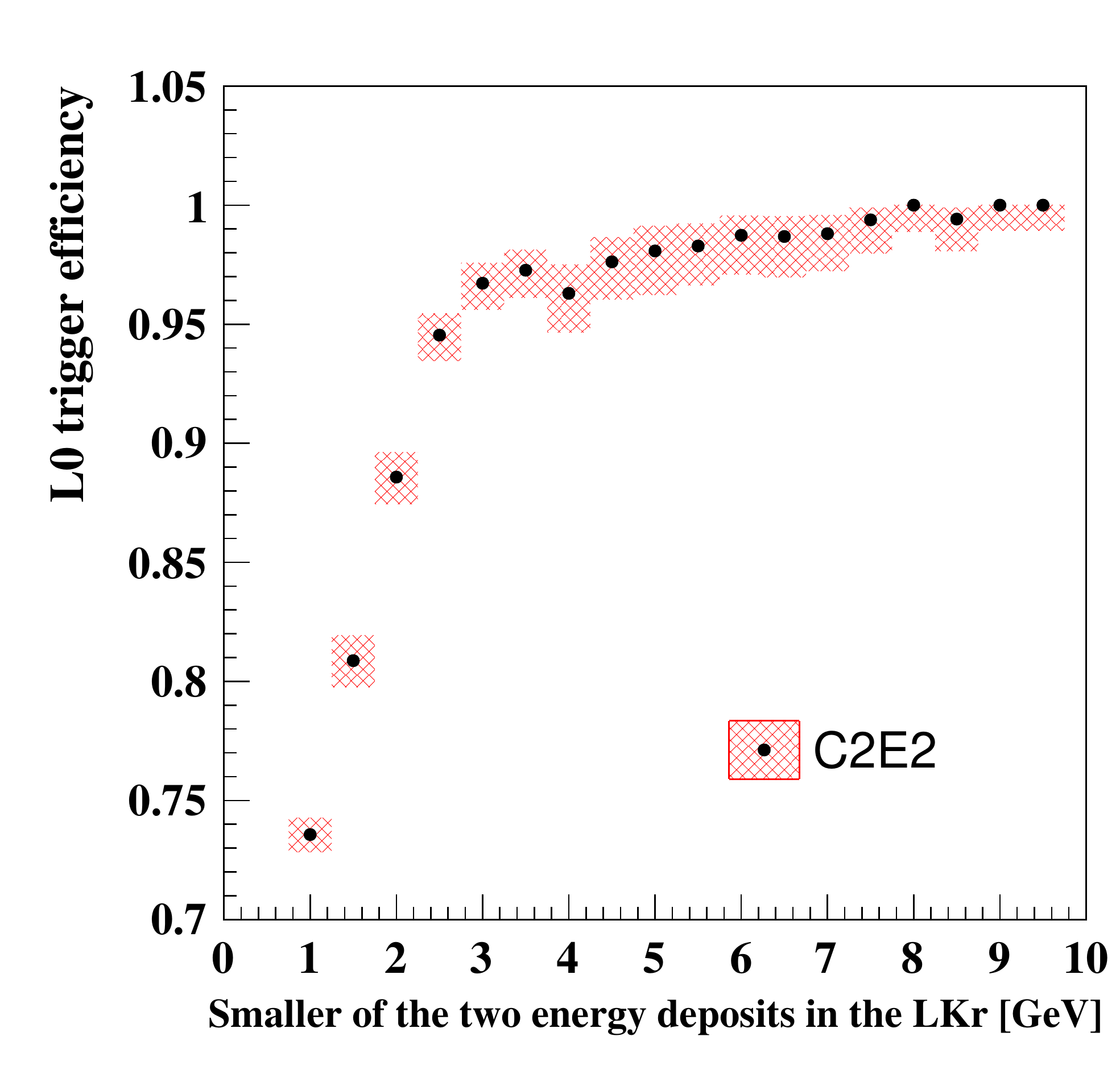}
 \caption{Efficiency of the C2E2 condition in beam-dump data, as a function of the smaller LKr energy deposit of the two tracks. The hatched areas indicate the statistical uncertainties of the measurements.}
 \label{fig:lkrprim}
\end{figure}

The efficiencies of the L0 trigger conditions Q2 and C2E2 are measured using data collected via the control trigger line (Table~\ref{table:dumpmode}), using a sample of events with two oppositely-charged tracks that form a vertex in the \FV.
The two tracks are required to extrapolate within the acceptance of the CHOD, NA48-CHOD, and LKr. When determining the efficiency of the Q2 trigger condition, the two tracks are required to extrapolate into different CHOD quadrants. When determining the efficiency of the C2E2 condition, at least \SI{1}{\giga\eV} in the LKr is required to be associated with each track, with no other LKr energy deposit.
The efficiency of the Q2 condition
evaluated as a function of the total momentum of the two tracks (total vertex momentum)
is larger than 99\% and does not vary as a function of the total vertex momentum (left panel of Fig.~\ref{fig:newchodprim}).
As the Q1 and Q2 conditions are correlated --  both are produced by the CHOD -- the efficiency is cross-checked using data collected via the neutral trigger line (right panel of Fig.~\ref{fig:newchodprim}). 
The efficiency is consistent between the two samples.
The efficiency of the C2E2 condition is evaluated as a function of the smaller energy deposit of the two tracks using a data sample collected via the charged trigger line (Table~\ref{table:dumpmode}).
The efficiency rises to 95\% at \SI{3}{\giga\eV}, and is close to 100\% at \SI{10}{\giga\eV} (Fig.~\ref{fig:lkrprim}).

\subsection{Simulation of the trigger response}
\begin{figure}[t]
\begin{subfigure}[b]{0.48\textwidth}
\includegraphics[width=\textwidth]{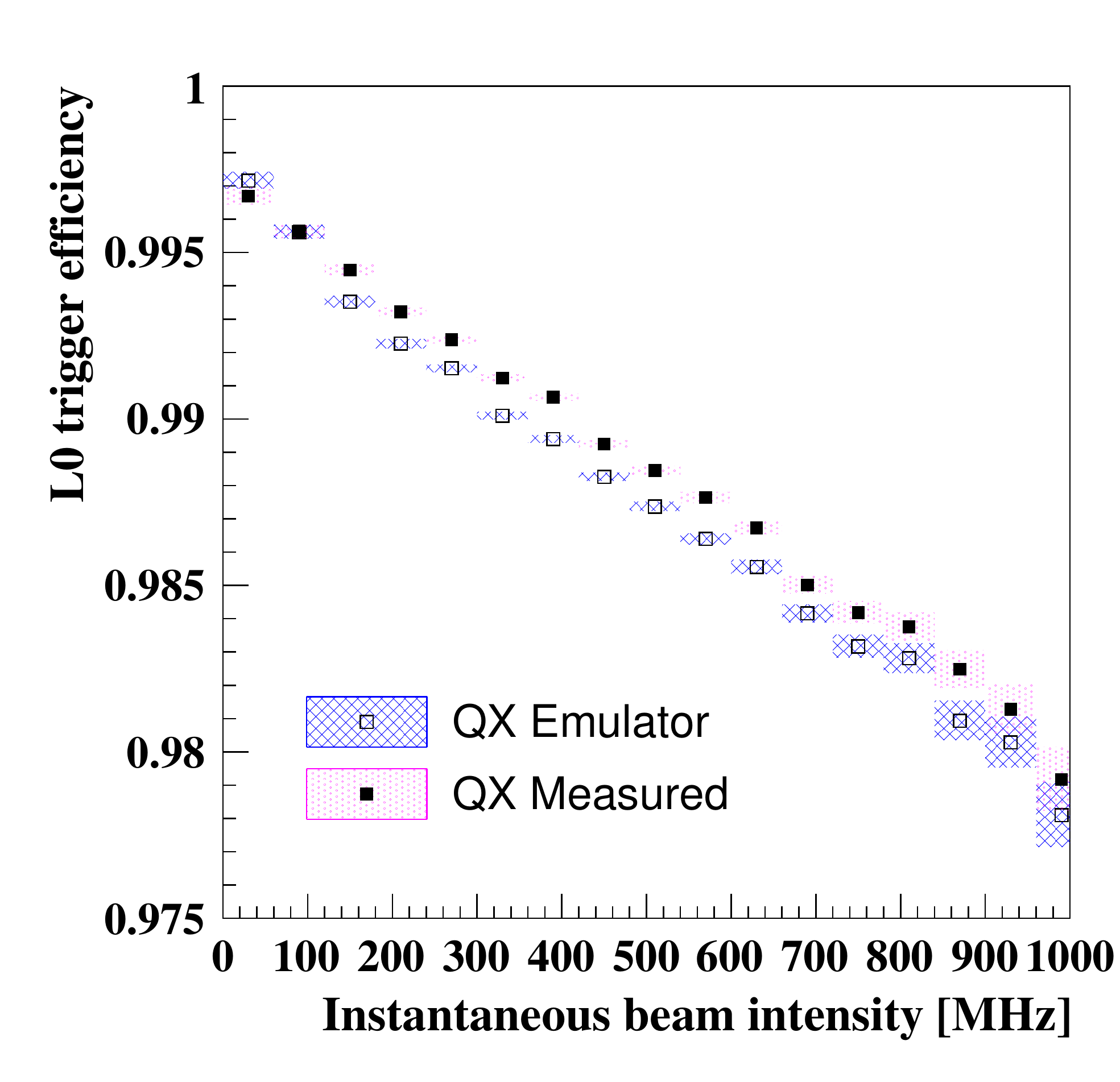}
\label{fig:K3pi-Emu-QX}
\end{subfigure}
\begin{subfigure}[b]{0.48\textwidth}
\includegraphics[width=\textwidth]{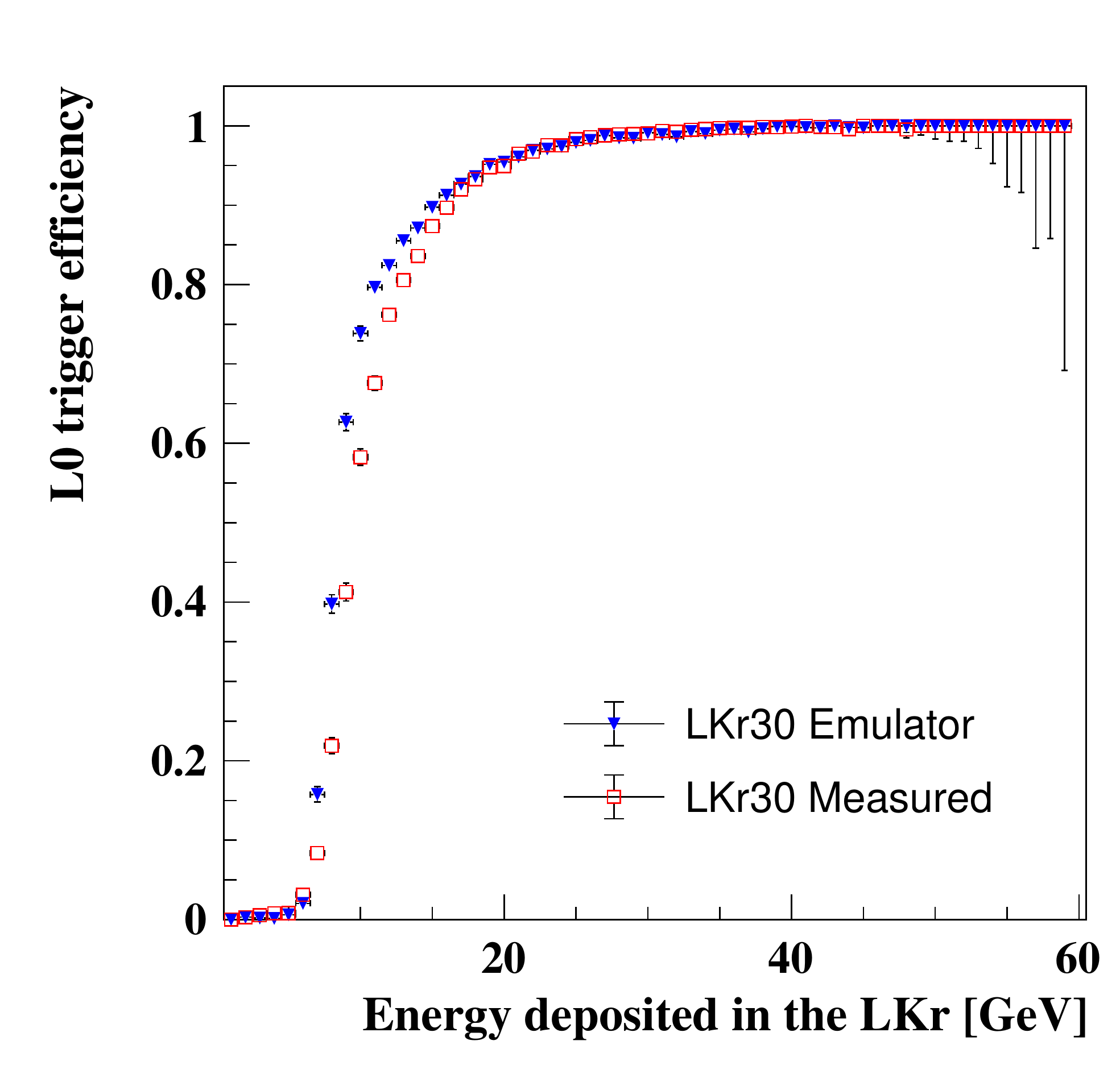}
\label{fig:K3pi-Emu-LKr30}
\end{subfigure}
\caption{Efficiencies of the L0 trigger conditions QX (left) and LKr30 (right) measured using \kpipipi\ events collected via the control trigger line, and simulated using the trigger emulator on the same data. The hatched areas indicate the statistical uncertainties of the measurements.}
\label{fig:K3pi-Emu}
\end{figure}

To simulate the L0 trigger response offline, a software emulation reproduces the firmware algorithms incorporated in the L0 trigger.
The L0 emulator output is compared to the measured trigger efficiency using a sample of \kpipipi\ decays obtained via the control trigger line. 
The QX efficiency, and its variation with instantaneous beam intensity, is reproduced to within 0.2\% (left panel of Fig.~\ref{fig:K3pi-Emu}).
The LKr30 efficiency is also reproduced to within 0.2\% for an energy deposit above \SI{20}{\giga\eV}~(right panel of Fig.~\ref{fig:K3pi-Emu}).
For an energy deposit below \SI{20}{\giga\eV}, the LKr30 efficiency is highly sensitive to the emulation of the cluster reconstruction, leading to a larger discrepancy between the emulated and measured efficiencies.
For the L1 trigger, part of the HLT software is integrated with the offline software, and as
the libraries and algorithms are shared between the HLT and offline software,
the L1 trigger response is reproduced exactly.

\section{Conclusions and outlook}
\label{sec:conclusion}

The NA62 trigger system was successfully operated during the 2016--2018 data-taking period, exhibiting good performance and reliability.
The flexibility of the system allowed its evolution during the data-taking, and several rate limitations have been identified and removed.

Two major developments are anticipated for the L0 trigger.
The first development is an upgraded version of the L0TP, installed for testing in 2021, that exploits newer hardware to handle a higher primitive rate and offer more flexibility in the conditions used to generate triggers.
The second development is a ring-finding algorithm for the RICH detector implemented 
on commercially-available Graphics Processing Units (GPU)~\cite{GPUs}. The \SI{1}{\milli\second} latency of the L0 trigger is sufficiently large for such a system to operate.

Improvements to the HLT software exploit advances in the \kpinunu\ analysis
to refine the trigger selection used to collect both the signal and the sample of ancillary data required for background estimation and systematic checks.
Several improvements have already been implemented, namely in the L1 STRAW algorithm to recover the efficiency loss at high \pip\ momentum, and in the L1 LAV algorithm to reduce the inefficiency caused by random veto.
Further exploitation of the available detector information can preserve specific categories of events that would be rejected by simpler algorithms, allowing a broader physics programme without impacting the \kpinunu\ trigger line.
One major improvement is to implement particle identification at L1 by matching tracks in the STRAW with Cherenkov rings in the RICH or energy deposits in the LKr.
The	latter is possible due to new hardware that transmits data from the \lzCalo\ to the L1 trigger.
An additional L2 trigger stage can also be used after the L1 to further reduce the data volume. Trigger algorithms implemented at L2 will have the advantage of being executed on complete events with information from all detectors.

\section*{Acknowledgements}
\input{acknow202209}


\end{document}

%% file: run1Trigperf_2.tex
\begin{center}
{\Large The NA62 Collaboration\renewcommand{\thefootnote}{\fnsymbol{footnote}}\footnotemark[1]\renewcommand{\thefootnote}{\arabic{footnote}}}\\
\end{center}
\begin{flushleft}
 E.~Cortina Gil\footnotemark[1],
 A.~Kleimenova\footnotemark[1]$^,$\renewcommand{\thefootnote}{\alphalph{\value{footnote}}}\footnotemark[1]\renewcommand{\thefootnote}{\arabic{footnote}},
E.~Minucci\footnotemark[1]$^,$\renewcommand{\thefootnote}{\alphalph{\value{footnote}}}\footnotemark[2]$^,$\footnotemark[3]\renewcommand{\thefootnote}{\arabic{footnote}},
 S.~Padolski\footnotemark[1]$^,$\renewcommand{\thefootnote}{\alphalph{\value{footnote}}}\footnotemark[4]\renewcommand{\thefootnote}{\arabic{footnote}},
 P.~Petrov\footnotemark[1],
 A.~Shaikhiev\footnotemark[1]$^,$\renewcommand{\thefootnote}{\alphalph{\value{footnote}}}\footnotemark[5]\renewcommand{\thefootnote}{\arabic{footnote}},
 R.~Volpe\footnotemark[1]$^,$\renewcommand{\thefootnote}{\alphalph{\value{footnote}}}\footnotemark[6]\renewcommand{\thefootnote}{\arabic{footnote}},  
 T.~Numao\footnotemark[2],
 Y.~Petrov\footnotemark[2],
 B.~Velghe\footnotemark[2],
 V.W.S.~Wong\footnotemark[2], 
 D.~Bryman\footnotemark[3]$^,$\renewcommand{\thefootnote}{\alphalph{\value{footnote}}}\footnotemark[7]\renewcommand{\thefootnote}{\arabic{footnote}},
 J.~Fu\footnotemark[3], 
 T.~Husek\footnotemark[4]$^,$\renewcommand{\thefootnote}{\alphalph{\value{footnote}}}\footnotemark[8]\renewcommand{\thefootnote}{\arabic{footnote}},
 J.~Jerhot\footnotemark[4]$^,$\renewcommand{\thefootnote}{\alphalph{\value{footnote}}}\footnotemark[9]\renewcommand{\thefootnote}{\arabic{footnote}},
 K.~Kampf,\footnotemark[4],
 M.~Zamkovsky\footnotemark[4]$^,$\renewcommand{\thefootnote}{\alphalph{\value{footnote}}}\footnotemark[2]\renewcommand{\thefootnote}{\arabic{footnote}},
 R.~Aliberti\footnotemark[5]$^,$\renewcommand{\thefootnote}{\alphalph{\value{footnote}}}\footnotemark[10]\renewcommand{\thefootnote}{\arabic{footnote}},
 G.~Khoriauli\footnotemark[5]$^,$\renewcommand{\thefootnote}{\alphalph{\value{footnote}}}\footnotemark[11]\renewcommand{\thefootnote}{\arabic{footnote}},
 J.~Kunze\footnotemark[5],
 D.~Lomidze\footnotemark[5]$^,$\renewcommand{\thefootnote}{\alphalph{\value{footnote}}}\footnotemark[12]\renewcommand{\thefootnote}{\arabic{footnote}},
 L.~Peruzzo\footnotemark[5],
 M.~Vormstein\footnotemark[5],
 R.~Wanke\footnotemark[5],   
 P.~Dalpiaz\footnotemark[6],
 M.~Fiorini\footnotemark[6],
 I.~Neri\footnotemark[6],
 A.~Norton\footnotemark[6]$^,$\renewcommand{\thefootnote}{\alphalph{\value{footnote}}}\footnotemark[13]\renewcommand{\thefootnote}{\arabic{footnote}},
 F.~Petrucci\footnotemark[6],
 H.~Wahl\footnotemark[6]$^,$\renewcommand{\thefootnote}{\alphalph{\value{footnote}}}\footnotemark[14]\renewcommand{\thefootnote}{\arabic{footnote}},  
 A.~Cotta Ramusino\footnotemark[7],
 A.~Gianoli\footnotemark[7],  
 E.~Iacopini\footnotemark[8],
 G.~Latino\footnotemark[8],
 M.~Lenti\footnotemark[8],
 A.~Parenti\footnotemark[8], 
 A.~Bizzeti\footnotemark[9]$^,$\renewcommand{\thefootnote}{\alphalph{\value{footnote}}}\footnotemark[15]\renewcommand{\thefootnote}{\arabic{footnote}}, 
 F.~Bucci\footnotemark[9], 
 A.~Antonelli\footnotemark[10],
G.~Georgiev\footnotemark[10]$^,$\renewcommand{\thefootnote}{\alphalph{\value{footnote}}}\footnotemark[16]\renewcommand{\thefootnote}{\arabic{footnote}}, 
V.~Kozhuharov\footnotemark[10]$^,$\renewcommand{\thefootnote}{\alphalph{\value{footnote}}}\footnotemark[16]\renewcommand{\thefootnote}{\arabic{footnote}},
G.~Lanfranchi\footnotemark[10],
S.~Martellotti\footnotemark[10],
M.~Moulson\footnotemark[10],
T.~Spadaro\footnotemark[10],
G.~Tinti\footnotemark[10], 
 F.~Ambrosino\footnotemark[11],
 T.~Capussela\footnotemark[11],
 M.~Corvino\footnotemark[11]$^,$\renewcommand{\thefootnote}{\alphalph{\value{footnote}}}\footnotemark[2]\renewcommand{\thefootnote}{\arabic{footnote}}, 
 D.~Di Filippo\footnotemark[11],
 R.~Fiorenza\footnotemark[11]$^,$\renewcommand{\thefootnote}{\alphalph{\value{footnote}}}\footnotemark[17]\renewcommand{\thefootnote}{\arabic{footnote}}, 
 P.~Massarotti\footnotemark[11],
 M.~Mirra\footnotemark[11],
 M.~Napolitano\footnotemark[11],
 G.~Saracino\footnotemark[11],  
 G.~Anzivino\footnotemark[12],
 F.~Brizioli\footnotemark[12]$^,$\renewcommand{\thefootnote}{\alphalph{\value{footnote}}}\footnotemark[2]\renewcommand{\thefootnote}{\arabic{footnote}}, 
 E.~Imbergamo\footnotemark[12],
 R.~Lollini\footnotemark[12],
 R.~Piandani\footnotemark[12]$^,$\renewcommand{\thefootnote}{\alphalph{\value{footnote}}}\footnotemark[18]\renewcommand{\thefootnote}{\arabic{footnote}}, 
 C.~Santoni\footnotemark[12],  
 M.~Barbanera\footnotemark[13],
 P.~Cenci\footnotemark[13],
 B.~Checcucci\footnotemark[13],
 P.~Lubrano\footnotemark[13],
 M.~Lupi\footnotemark[13]$^,$\renewcommand{\thefootnote}{\alphalph{\value{footnote}}}\footnotemark[19]\renewcommand{\thefootnote}{\arabic{footnote}}, 
 M.~Pepe\footnotemark[13],
 M.~Piccini\footnotemark[13],  
{
 F.~Costantini\footnotemark[14],
L.~Di Lella\footnotemark[14]$^,$\renewcommand{\thefootnote}{\alphalph{\value{footnote}}}\footnotemark[14]\renewcommand{\thefootnote}{\arabic{footnote}}, 
 N.~Doble\footnotemark[14]$^,$\renewcommand{\thefootnote}{\alphalph{\value{footnote}}}\footnotemark[14]\renewcommand{\thefootnote}{\arabic{footnote}}, 
 M.~Giorgi\footnotemark[14],
 S.~Giudici\footnotemark[14],
 G.~Lamanna\footnotemark[14],
 E.~Lari\footnotemark[14],
 E.~Pedreschi\footnotemark[14],
 M.~Sozzi\footnotemark[14],  
 C.~Cerri\footnotemark[15],
 R.~Fantechi\footnotemark[15],
 L.~Pontisso\footnotemark[15]$^,$\renewcommand{\thefootnote}{\alphalph{\value{footnote}}}\footnotemark[20]\renewcommand{\thefootnote}{\arabic{footnote}}, 
 F.~Spinella\footnotemark[15],  
 I.~Mannelli\footnotemark[16],   
 G.~D'Agostini\footnotemark[17], 
 M.~Raggi\footnotemark[17],  
 A.~Biagioni\footnotemark[18], 
 P.~Cretaro\footnotemark[18], 
 O.~Frezza\footnotemark[18], 
 E.~Leonardi\footnotemark[18], 
 A.~Lonardo\footnotemark[18], 
 M.~Turisini\footnotemark[18], 
 P.~Valente\footnotemark[18], 
 P.~Vicini\footnotemark[18],  
 R.~Ammendola\footnotemark[19], 
 V.~Bonaiuto\footnotemark[19]$^,$\renewcommand{\thefootnote}{\alphalph{\value{footnote}}}\footnotemark[21]\renewcommand{\thefootnote}{\arabic{footnote}}, 
 A.~Fucci\footnotemark[19], 
 A.~Salamon\footnotemark[19], 
 F.~Sargeni\footnotemark[19]$^,$\renewcommand{\thefootnote}{\alphalph{\value{footnote}}}\footnotemark[22]\renewcommand{\thefootnote}{\arabic{footnote}},   
 R.~Arcidiacono\footnotemark[20]$^,$\renewcommand{\thefootnote}{\alphalph{\value{footnote}}}\footnotemark[23]\renewcommand{\thefootnote}{\arabic{footnote}}, 
 B.~Bloch-Devaux\footnotemark[20],
 M.~Boretto\footnotemark[20]$^,$\renewcommand{\thefootnote}{\alphalph{\value{footnote}}}\footnotemark[2]\renewcommand{\thefootnote}{\arabic{footnote}}, 
 E.~Menichetti\footnotemark[20],
 E.~Migliore\footnotemark[20],
 D.~Soldi\footnotemark[20],   
 C.~Biino\footnotemark[21],
 A.~Filippi\footnotemark[21],
 F.~Marchetto\footnotemark[21],  
 J.~Engelfried\footnotemark[22],
 N.~Estrada-Tristan\footnotemark[22]$^,$\renewcommand{\thefootnote}{\alphalph{\value{footnote}}}\footnotemark[24]\renewcommand{\thefootnote}{\arabic{footnote}},  
 A.M.~Bragadireanu\footnotemark[23],
 S.A.~Ghinescu\footnotemark[23],
 O.E.~Hutanu\footnotemark[23], 
 A.~Baeva\footnotemark[24],
 D.~Baigarashev\footnotemark[24]$^,$\renewcommand{\thefootnote}{\alphalph{\value{footnote}}}\footnotemark[25]\renewcommand{\thefootnote}{\arabic{footnote}}, 
 D.~Emelyanov\footnotemark[24],
 T.~Enik\footnotemark[24],
 V.~Falaleev\footnotemark[24]$^,$\renewcommand{\thefootnote}{\alphalph{\value{footnote}}}\footnotemark[26]\renewcommand{\thefootnote}{\arabic{footnote}}, 
 V.~Kekelidze\footnotemark[24],
 A.~Korotkova\footnotemark[24],
 L.~Litov\footnotemark[24]$^,$\renewcommand{\thefootnote}{\alphalph{\value{footnote}}}\footnotemark[16]\renewcommand{\thefootnote}{\arabic{footnote}}, 
 D.~Madigozhin\footnotemark[24],
 M.~Misheva\footnotemark[24]$^,$\renewcommand{\thefootnote}{\alphalph{\value{footnote}}}\footnotemark[27]\renewcommand{\thefootnote}{\arabic{footnote}}, 
 N.~Molokanova\footnotemark[24],
 S.~Movchan\footnotemark[24],
 I.~Polenkevich\footnotemark[24],
 Yu.~Potrebenikov\footnotemark[24],
 S.~Shkarovskiy\footnotemark[24],
 A.~Zinchenko\footnotemark[24]$^,$\renewcommand{\thefootnote}{\fnsymbol{footnote}}\footnotemark[2]\renewcommand{\thefootnote}{\arabic{footnote}},  
 S.~Fedotov\footnotemark[25],
 E.~Gushchin\footnotemark[25],
 A.~Khotyantsev\footnotemark[25],
 Y.~Kudenko\footnotemark[25]$^,$\renewcommand{\thefootnote}{\alphalph{\value{footnote}}}\footnotemark[28]\renewcommand{\thefootnote}{\arabic{footnote}},  
 V.~Kurochka\footnotemark[25],
 M.~Medvedeva\footnotemark[25],
 A.~Mefodev \footnotemark[25], 
 S.~Kholodenko\footnotemark[26],
 V.~Kurshetsov\footnotemark[26],
 V.~Obraztsov\footnotemark[26],
 A.~Ostankov\footnotemark[26]$^,$\renewcommand{\thefootnote}{\fnsymbol{footnote}}\footnotemark[2]\renewcommand{\thefootnote}{\arabic{footnote}},
 V.~Semenov\footnotemark[26]$^,$\renewcommand{\thefootnote}{\fnsymbol{footnote}}\footnotemark[2]\renewcommand{\thefootnote}{\arabic{footnote}},
 V.~Sugonyaev\footnotemark[26],
 O.~Yushchenko\footnotemark[26],  
 L.~Bician\footnotemark[27]$^,$\renewcommand{\thefootnote}{\alphalph{\value{footnote}}}\footnotemark[29]\renewcommand{\thefootnote}{\arabic{footnote}},  
 T.~Blazek\footnotemark[27],
 V.~Cerny\footnotemark[27],
 Z.~Kucerova\footnotemark[27],  
 J.~Bernhard\footnotemark[28],
 A.~Ceccucci\footnotemark[28],
 H.~Danielsson\footnotemark[28],
 N.~De Simone\footnotemark[28]$^,$\renewcommand{\thefootnote}{\alphalph{\value{footnote}}}\footnotemark[30]\renewcommand{\thefootnote}{\arabic{footnote}},
 F.~Duval\footnotemark[28],
 B.~D\"obrich\footnotemark[28],
 L.~Federici\footnotemark[28],
 E.~Gamberini\footnotemark[28],
 L.~Gatignon\footnotemark[28]$^,$\renewcommand{\thefootnote}{\alphalph{\value{footnote}}}\footnotemark[31]\renewcommand{\thefootnote}{\arabic{footnote}},
 R.~Guida\footnotemark[28],
 F.~Hahn\footnotemark[28]$^,$\renewcommand{\thefootnote}{\fnsymbol{footnote}}\footnotemark[2]\renewcommand{\thefootnote}{\arabic{footnote}},
 E.B.~Holzer\footnotemark[28],
 B.~Jenninger\footnotemark[28],
 M.~Koval\footnotemark[28]$^,$\renewcommand{\thefootnote}{\alphalph{\value{footnote}}}\footnotemark[29]\renewcommand{\thefootnote}{\arabic{footnote}},
 P.~Laycock\footnotemark[28]$^,$\renewcommand{\thefootnote}{\alphalph{\value{footnote}}}\footnotemark[4]\renewcommand{\thefootnote}{\arabic{footnote}},
 G.~Lehmann Miotto\footnotemark[28],
 P.~Lichard\footnotemark[28],
 A.~Mapelli\footnotemark[28],
 R.~Marchevski\footnotemark[28]$^,$\renewcommand{\thefootnote}{\alphalph{\value{footnote}}}\footnotemark[32]\renewcommand{\thefootnote}{\arabic{footnote}},
 K.~Massri\footnotemark[28],
 M.~Noy\footnotemark[28],
 V.~Palladino\footnotemark[28],
 M.~Perrin-Terrin\footnotemark[28]$^,$\renewcommand{\thefootnote}{\alphalph{\value{footnote}}}\footnotemark[33]$^,$\footnotemark[34]\renewcommand{\thefootnote}{\arabic{footnote}},
 J.~Pinzino\footnotemark[28]$^,$\renewcommand{\thefootnote}{\alphalph{\value{footnote}}}\footnotemark[35]\renewcommand{\thefootnote}{\arabic{footnote}}\renewcommand{\thefootnote}{\arabic{footnote}},
 V.~Ryjov\footnotemark[28],
 S.~Schuchmann\footnotemark[28],
 S.~Venditti\footnotemark[28],  
 T.~Bache\footnotemark[29],
 M.B.~Brunetti\footnotemark[29]$^,$\renewcommand{\thefootnote}{\alphalph{\value{footnote}}}\footnotemark[36]\renewcommand{\thefootnote}{\arabic{footnote}},
 V.~Duk\footnotemark[29]$^,$\renewcommand{\thefootnote}{\alphalph{\value{footnote}}}\footnotemark[6]\renewcommand{\thefootnote}{\arabic{footnote}},
 V.~Fascianelli\footnotemark[29]$^,$\renewcommand{\thefootnote}{\alphalph{\value{footnote}}}\footnotemark[37]\renewcommand{\thefootnote}{\arabic{footnote}},
 J. R.~Fry\footnotemark[29],
 F.~Gonnella\footnotemark[29],
 E.~Goudzovski\footnotemark[29],
 J.~Henshaw\footnotemark[29],
 L.~Iacobuzio\footnotemark[29],
 C.~Lazzeroni\footnotemark[29],
 N.~Lurkin\footnotemark[29]$^,$\renewcommand{\thefootnote}{\alphalph{\value{footnote}}}\footnotemark[9]\renewcommand{\thefootnote}{\arabic{footnote}},
 F.~Newson\footnotemark[29],
 C.~Parkinson\footnotemark[29]$^,$\renewcommand{\thefootnote}{\fnsymbol{footnote}}\footnotemark[1]\renewcommand{\thefootnote}{\arabic{footnote}},
 A.~Romano\footnotemark[29]$^,$\renewcommand{\thefootnote}{\fnsymbol{footnote}}\footnotemark[1]\renewcommand{\thefootnote}{\arabic{footnote}},
 A.~Sergi\footnotemark[29]$^,$\renewcommand{\thefootnote}{\alphalph{\value{footnote}}}\footnotemark[38]\renewcommand{\thefootnote}{\arabic{footnote}},
 A.~Sturgess\footnotemark[29],
 J.~Swallow\footnotemark[29]$^,$\renewcommand{\thefootnote}{\alphalph{\value{footnote}}}\footnotemark[2]\renewcommand{\thefootnote}{\arabic{footnote}},
 A.~Tomczak\footnotemark[29],  
 H.~Heath\footnotemark[30],
 R.~Page\footnotemark[30],
 S.~Trilov\footnotemark[30], 
 B.~Angelucci\footnotemark[31],
 D.~Britton\footnotemark[31],
 C.~Graham\footnotemark[31],
 D.~Protopopescu\footnotemark[31],  
 J.~Carmignani\footnotemark[32]$^,$\renewcommand{\thefootnote}{\alphalph{\value{footnote}}}\footnotemark[39]\renewcommand{\thefootnote}{\arabic{footnote}}, 
 J.B.~Dainton\footnotemark[32],
 R.W. L.~Jones\footnotemark[32],
 G.~Ruggiero\footnotemark[32]$^,$\renewcommand{\thefootnote}{\alphalph{\value{footnote}}}\footnotemark[40]\renewcommand{\thefootnote}{\arabic{footnote}},   
 L.~Fulton\footnotemark[33],
 D.~Hutchcroft\footnotemark[33],
 E.~Maurice\footnotemark[33]$^,$\renewcommand{\thefootnote}{\alphalph{\value{footnote}}}\footnotemark[41]\renewcommand{\thefootnote}{\arabic{footnote}},  
 B.~Wrona\footnotemark[33],  
 A.~Conovaloff\footnotemark[34],
 P.~Cooper\footnotemark[34],
 D.~Coward\footnotemark[34]$^,$\renewcommand{\thefootnote}{\alphalph{\value{footnote}}}\footnotemark[42]\renewcommand{\thefootnote}{\arabic{footnote}},  
 P.~Rubin\footnotemark[34]  
}\end{flushleft}
\newlength{\basefootnotesep}
\setlength{\basefootnotesep}{\footnotesep}
\begin{flushleft}
\setcounter{footnote}{0}
\renewcommand{\thefootnote}{\fnsymbol{footnote}}
\footnotetext[1]{Corresponding authors:  A.~Romano, C.~Parkinson, email: angela.romano@cern.ch, chris.parkinson@cern.ch}
\footnotetext[2]{Deceased}
\renewcommand{\thefootnote}{\arabic{footnote}}
$^{1}$ 
Universit\'e Catholique de Louvain, B-1348 Louvain-La-Neuve, Belgium \\
$^{2}$ 
TRIUMF, Vancouver, British Columbia, V6T 2A3, Canada \\
$^{3}$
University of British Columbia, Vancouver, British Columbia, V6T 1Z4, Canada \\
$^{4}$
Charles University, 116 36 Prague 1, Czech Republic \\
$^{5}$
Institut f\"ur Physik and PRISMA Cluster of Excellence, Universit\"at Mainz, D-55099 Mainz, Germany \\
$^{6}$
Dipartimento di Fisica e Scienze della Terra dell'Universit\`a e INFN, Sezione di Ferrara, I-44122 Ferrara, Italy \\
$^{7}$
INFN, Sezione di Ferrara, I-44122 Ferrara, Italy \\
$^{8}$
Dipartimento di Fisica e Astronomia dell'Universit\`a e INFN, Sezione di Firenze, I-50019 Sesto Fiorentino, Italy \\
$^{9}$
INFN, Sezione di Firenze, I-50019 Sesto Fiorentino, Italy \\
$^{10}$
Laboratori Nazionali di Frascati, I-00044 Frascati, Italy \\
$^{11}$
Dipartimento di Fisica ``Ettore Pancini'' e INFN, Sezione di Napoli, I-80126 Napoli, Italy \\
$^{12}$
Dipartimento di Fisica e Geologia dell'Universit\`a e INFN, Sezione di Perugia, I-06100 Perugia, Italy \\
$^{13}$
INFN, Sezione di Perugia, I-06100 Perugia, Italy \\
$^{14}$
Dipartimento di Fisica dell'Universit\`a e INFN, Sezione di Pisa, I-56100 Pisa, Italy \\
$^{15}$
INFN, Sezione di Pisa, I-56100 Pisa, Italy \\
$^{16}$
Scuola Normale Superiore e INFN, Sezione di Pisa, I-56100 Pisa, Italy \\
$^{17}$
Dipartimento di Fisica, Sapienza Universit\`a di Roma e INFN, Sezione di Roma I, I-00185 Roma, Italy \\
$^{18}$
INFN, Sezione di Roma I, I-00185 Roma, Italy \\
$^{19}$
INFN, Sezione di Roma Tor Vergata, I-00133 Roma, Italy \\
$^{20}$
Dipartimento di Fisica dell'Universit\`a e INFN, Sezione di Torino, I-10125 Torino, Italy \\
$^{21}$
INFN, Sezione di Torino, I-10125 Torino, Italy \\
$^{22}$
Instituto de F\'isica, Universidad Aut\'onoma de San Luis Potos\'i, 78240 San Luis Potos\'i, Mexico \\
$^{23}$
Horia Hulubei National Institute for R\&D in Physics and Nuclear Engineering, 077125 Bucharest-Magurele, Romania \\
$^{24}$
Joint Institute for Nuclear Research, 141980 Dubna (MO), Russia \\
$^{25}$
Institute for Nuclear Research of the Russian Academy of Sciences, 117312 Moscow, Russia \\
$^{26}$
Institute for High Energy Physics of the Russian Federation, State Research Center ``Kurchatov Institute", 142281 Protvino (MO), Russia \\
$^{27}$
Faculty of Mathematics, Physics and Informatics, Comenius University, 842 48, Bratislava, Slovakia \\
$^{28}$
CERN,  European Organization for Nuclear Research, CH-1211 Geneva 23, Switzerland \\
$^{29}$
School of Physics and Astronomy, University of Birmingham, Birmingham, B15 2TT, UK \\
$^{30}$
School of Physics, University of Bristol, Bristol, BS8 1TH, UK \\
$^{31}$
School of Physics and Astronomy, University of Glasgow, Glasgow, G12 8QQ, UK \\
$^{32}$
Faculty of Science and Technology, University of Lancaster, Lancaster, LA1 4YW, UK \\
$^{33}$
School of Physical Sciences, University of Liverpool, Liverpool, L69 7ZE, UK \\
$^{34}$
Physics and Astronomy Department, George Mason University, Fairfax, VA 22030, USA
\end{flushleft}
\renewcommand{\thefootnote}{\alphalph{\value{footnote}}}
\footnotesize
$^{\alphalph{1}}$Present address: Faculty of Mathematics, Physics and Informatics, Comenius University, 842 48, Bratislava, Slovakia \\
$^{\alphalph{2}}$Present address: CERN,  European Organization for Nuclear Research, CH-1211 Geneva 23, Switzerland \\
$^{\alphalph{3}}$Also at Laboratori Nazionali di Frascati, I-00044 Frascati, Italy \\
$^{\alphalph{4}}$Present address: Brookhaven National Laboratory, Upton, NY 11973, USA \\
$^{\alphalph{5}}$Present address: School of Physics and Astronomy, University of Birmingham, Birmingham, B15 2TT, UK \\
$^{\alphalph{6}}$Present address: INFN, Sezione di Perugia, I-06100 Perugia, Italy \\
$^{\alphalph{7}}$Also at TRIUMF, Vancouver, British Columbia, V6T 2A3, Canada \\
$^{\alphalph{8}}$Present address: Department of Astronomy and Theoretical Physics, Lund University, Lund, SE 223-62, Sweden \\
$^{\alphalph{9}}$Present address: Universit\'e Catholique de Louvain, B-1348 Louvain-La-Neuve, Belgium \\
$^{\alphalph{10}}$Present address: Institut f\"ur Kernphysik and Helmholtz Institute Mainz, Universit\"at Mainz, Mainz, D-55099, Germany \\
$^{\alphalph{11}}$Present address: Universit\"at W\"urzburg, D-97070 W\"urzburg, Germany \\
$^{\alphalph{12}}$Present address: European XFEL GmbH, D-22761 Hamburg, Germany \\
$^{\alphalph{13}}$Present address: University of Glasgow, Glasgow, G12 8QQ, UK \\
$^{\alphalph{14}}$Present address: Institut f\"ur Physik and PRISMA Cluster of Excellence, Universit\"at Mainz, D-55099 Mainz, Germany \\
$^{\alphalph{15}}$Also at Dipartimento di Scienze Fisiche, Informatiche e Matematiche, Universit\`a di Modena e Reggio Emilia, I-41125 Modena, Italy \\
$^{\alphalph{16}}$Also at Faculty of Physics, University of Sofia, BG-1164 Sofia, Bulgaria \\
$^{\alphalph{17}}$Present address: Scuola Superiore Meridionale e INFN, Sezione di Napoli, I-80138 Napoli, Italy \\
$^{\alphalph{18}}$Present address: Instituto de F\'isica, Universidad Aut\'onoma de San Luis Potos\'i, 78240 San Luis Potos\'i, Mexico \\
$^{\alphalph{19}}$Present address: Institut am Fachbereich Informatik und Mathematik, Goethe Universit\"at, D-60323 Frankfurt am Main, Germany \\
$^{\alphalph{20}}$Present address: INFN, Sezione di Roma I, I-00185 Roma, Italy \\
$^{\alphalph{21}}$Also at Department of Industrial Engineering, University of Roma Tor Vergata, I-00173 Roma, Italy \\
$^{\alphalph{22}}$Also at Department of Electronic Engineering, University of Roma Tor Vergata, I-00173 Roma, Italy \\
$^{\alphalph{23}}$Also at Universit\`a degli Studi del Piemonte Orientale, I-13100 Vercelli, Italy \\
$^{\alphalph{24}}$Also at Universidad de Guanajuato, 36000 Guanajuato, Mexico \\
$^{\alphalph{25}}$Also at L.N. Gumilyov Eurasian National University, 010000 Nur-Sultan, Kazakhstan \\
$^{\alphalph{26}}$Also at Institute for Nuclear Research of the Russian Academy of Sciences, 117312 Moscow, Russia \\
$^{\alphalph{27}}$Present address: Institute of Nuclear Research and Nuclear Energy of Bulgarian Academy of Science (INRNE-BAS), BG-1784 Sofia, Bulgaria \\
$^{\alphalph{28}}$Also at National Research Nuclear University (MEPhI), 115409 Moscow and Moscow Institute of Physics and Technology, 141701 Moscow region, Moscow, Russia \\
$^{\alphalph{29}}$Present address: Charles University, 116 36 Prague 1, Czech Republic \\
$^{\alphalph{30}}$Present address: DESY, D-15738 Zeuthen, Germany \\
$^{\alphalph{31}}$Present address: University of Lancaster, Lancaster, LA1 4YW, UK \\
$^{\alphalph{32}}$Present address: Weizmann Institute, Rehovot, 76100, Israel \\
$^{\alphalph{33}}$Present address: Aix Marseille University, CNRS/IN2P3, CPPM, F-13288, Marseille, France \\
$^{\alphalph{34}}$Also at Universit\'e Catholique de Louvain, B-1348 Louvain-La-Neuve, Belgium \\
$^{\alphalph{35}}$Present address: INFN, Sezione di Pisa, I-56100 Pisa, Italy \\
$^{\alphalph{36}}$Present address: Department of Physics, University of Warwick, Coventry, CV4 7AL, UK \\
$^{\alphalph{37}}$Present address: Center for theoretical neuroscience, Columbia University, New York, NY 10027, USA \\
$^{\alphalph{38}}$Present address: Dipartimento di Fisica dell'Universit\`a e INFN, Sezione di Genova, I-16146 Genova, Italy \\
$^{\alphalph{39}}$Present address: University of Liverpool, Liverpool, L69 7ZE, UK \\
$^{\alphalph{40}}$Present address: Dipartimento di Fisica e Astronomia dell'Universit\`a e INFN, Sezione di Firenze, I-50019 Sesto Fiorentino, Italy \\
$^{\alphalph{41}}$Present address: Laboratoire Leprince Ringuet, F-91120 Palaiseau, France \\
$^{\alphalph{42}}$Also at SLAC National Accelerator Laboratory, Stanford University, Menlo Park, CA 94025, USA
\normalsize
\clearpage

%% file: acknow202209.tex
It is a pleasure to express our appreciation to the staff of the CERN laboratory and the technical
staff of the participating laboratories and universities for their efforts in the operation of the
experiment and data processing.

The cost of the experiment and its auxiliary systems was supported by the funding agencies of 
the Collaboration Institutes. We are particularly indebted to: 
F.R.S.-FNRS (Fonds de la Recherche Scientifique - FNRS), under Grants No. 4.4512.10, 1.B.258.20, Belgium;
CECI (Consortium des Equipements de Calcul Intensif), funded by the Fonds de la Recherche Scientifique de Belgique (F.R.S.-FNRS) under Grant No. 2.5020.11 and by the Walloon Region, Belgium;
NSERC (Natural Sciences and Engineering Research Council), funding SAPPJ-2018-0017,  Canada;
MEYS (Ministry of Education, Youth and Sports) funding LM 2018104, Czech Republic;
BMBF (Bundesministerium f\"{u}r Bildung und Forschung) contracts 05H12UM5, 05H15UMCNA and 05H18UMCNA, Germany;
INFN  (Istituto Nazionale di Fisica Nucleare),  Italy;
MIUR (Ministero dell'Istruzione, dell'Universit\`a e della Ricerca),  Italy;
CONACyT  (Consejo Nacional de Ciencia y Tecnolog\'{i}a),  Mexico;
IFA (Institute of Atomic Physics) Romanian 
CERN-RO No. 1/16.03.2016 
and Nucleus Programme PN 19 06 01 04,  Romania;
INR-RAS (Institute for Nuclear Research of the Russian Academy of Sciences), Moscow, Russia; 
JINR (Joint Institute for Nuclear Research), Dubna, Russia; 
NRC (National Research Center)  ``Kurchatov Institute'' and MESRF (Ministry of Education and Science of the Russian Federation), Russia; 
MESRS  (Ministry of Education, Science, Research and Sport), Slovakia; 
CERN (European Organization for Nuclear Research), Switzerland; 
STFC (Science and Technology Facilities Council), United Kingdom;
NSF (National Science Foundation) Award Numbers 1506088 and 1806430,  U.S.A.;
ERC (European Research Council)  ``UniversaLepto'' advanced grant 268062, ``KaonLepton'' starting grant 336581, Europe.

Individuals have received support from:
Charles University Research Center (UNCE/SCI/ 013), Czech Republic;
Ministero dell'Istruzione, dell'Universit\`a e della Ricerca (MIUR  ``Futuro in ricerca 2012''  grant RBFR12JF2Z, Project GAP), Italy;
Russian Science Foundation (RSF 19-72-10096), Russia;
the Royal Society  (grants UF100308, UF0758946), United Kingdom;
STFC (Rutherford fellowships ST/J00412X/1, ST/M005798/1), United Kingdom;
ERC (grants 268062,  336581 and  starting grant 802836 ``AxScale'');
EU Horizon 2020 (Marie Sk\l{}odowska-Curie grants 701386, 754496, 842407, 893101, 101023808).

The data used in this paper were collected before February 2022.